\documentclass[onecolumn,noshowpacs,nofootinbib,colorlinks,hyperindex,unicode,11pt]{revtex4}
\usepackage{graphicx}
\usepackage{tikz}

\usepackage[compat=1.1.0]{tikz-feynman}

\usepackage{amsfonts,amsmath,amssymb,tikz,accents}
\usepackage[eulergreek]{sansmath}
\usepackage{indentfirst,bigints}
\usepackage{hyperref,upgreek}
\usepackage{bm,graphics,color}
\usepackage{feynmp-auto}
\usepackage{slashed}
\counterwithout{equation}{section}
\usepackage{hyperref}

\hypersetup{hidelinks,backref=true,pagebackref=true,hyperindex=true,colorlinks=true,breaklinks=true,urlcolor= blue}
\hypersetup{%
  colorlinks = true,
  linkcolor  = blue,
  citecolor = cyan,
}
\usepackage{textgreek}
\usepackage{color,xcolor}
\newcommand{\gdualn}[1]{\overset{\:{}^{{}^{\boldsymbol{\neg}}}}{\smash[t]{#1}}} 

\def\0{\mbox{\boldmath$\displaystyle\mathbb{O}$}}

\def\I{\openone}
\def\openone{\mathbb I}
\def\s{\mbox{\boldmath$\displaystyle\boldsymbol{\sigma}$}}

\def\p{\mbox{{{\scalebox{0.9}{$\displaystyle\bf{p}$}}}}}

\newcommand\orcidroldao{{\href{https://orcid.org/0000-0003-3978-532X}{\orcidicon}}}
\newcommand{\orcidicon}{%
	\begin{tikzpicture}
	\draw[lime, fill=lime] (0,0)
		circle [radius=0.16]
		node[white] {{\fontfamily{qag}\selectfont \tiny ID}};
	\draw[white, fill=white] (-0.0625,0.095)
		circle [radius=0.007];
	\end{tikzpicture}	\hspace{-2mm}
}
\newcommand\orcidg{{\href{https://orcid.org/0000-0002-7942-7941}{\orcidicon}}}
\newcommand\orcidNog{{\href{https://orcid.org/0000-0002-1827-1031}{\orcidicon}}}

\usepackage{float,xcolor,upgreek}
\usepackage{color} 
\usepackage{tikz-feynman}
\tikzfeynmanset{compat=1.0.0}

\newcommand{\beq}{\begin{eqnarray}}
\newcommand{\eeq}{\end{eqnarray}}
\newcommand{\bea}{\begin{eqnarray}}
\newcommand{\eea}{\end{eqnarray}}

\begin{document}

\title{Fermionic dark matter-photon quantum interaction: A mechanism for darkness}

\author{G. B. de Gracia\orcidg{}}
\affiliation{Federal University of ABC, Center of Mathematics,  Santo Andr\'e, 09210-580, Brazil.}
\email{g.gracia@ufabc.edu.br}
\author{A. A. Nogueira\orcidNog{}}
\affiliation{Federal University of Alfenas, Physics Department: Institute of Exact Sciences, Alfenas 37133-840, Brazil.}
\email{anderson.nogueira@unifal-mg.edu.br}
\author{R. da Rocha\orcidroldao\!\!}
\affiliation{Federal University of ABC, Center of Mathematics,  Santo Andr\'e 09210-580, Brazil.}
\email{roldao.rocha@ufabc.edu.br}

\begin{abstract}

\textcolor{black}{Mass dimension one} fermionic fields are prime candidates to describe dark matter, due to their intrinsic neutral nature, as they are constructed as eigenstates of the charge conjugation operator with dual helicity. To formulate the meaning of the darkness, the fermion-photon coupling is scrutinized with a Pauli-like interaction, and the path integral is then formulated from the phase space constraint structure.  Ward--Takahashi-like identities and Schwinger--Dyson equations,  together with renormalizability, are employed  to investigate a phenomenological mechanism to avoid external light signals. Accordingly, the non-polarized pair annihilation and Compton-like processes are shown to vanish at the limit of small scattering angles even if considering 1-loop radiative corrections, reinforcing the dark matter interpretation. 
However, dark matter interactions with nucleons are still possible. Motivated by recent  nucleon-recoil experiments to detect dark matter, we furnish a consistent theoretical setup to describe interaction with the photon compatible with the prevalence of darkness.

\end{abstract}
\maketitle
\section{Introduction}


Dark matter (DM) comprises the prevalent  gravitational mass in the universe, although the very nature of the DM sector remains undisclosed. However, it is evidenced by indirect observations. DM nature has observed effects only in the gravitational sector, demanding  other
 eventual interactions with Standard Model elementary particles, if any, to be extremely weak. 
 Interactions between DM and the photon comprise great scientific interest, as the photon is the core carrier of cosmological and astrophysical probes. 
Spin-1/2  mass dimension one fermionic quantum  fields  are prime candidates to describe DM, since they are, by construction, neutral fermions under gauge interactions, therefore implementing darkness \cite{Ahluwalia:2022ttu}. ELKO\footnote{ELKO is an acronym for the naming ``eigenspinors of the charge conjugation operator'' (Eigenspinoren des
Ladungskonjugationsoperators, in German)
\cite{Ahluwalia:2004ab}.} is 
a prominent archetype of \textcolor{black}{the spinor expansion coefficients for the } mass 
dimension one spin one-half fields and 
permits  unsuppressed  
tree-level couplings with 
photons, Higgs boson pairs, 
and self-interaction terms, 
which are invariant under 
gauge transformations. 
Mass dimension one fields are candidates for describing DM, as  effective couplings are  suppressed to almost
the entire Standard Model spectrum, except for the fermion-photon tree-level interaction, which is implemented by a dimensionless coupling. This feature is corroborated by experimental limits on photon-DM interactions. Although   interactions with the Higgs boson contribute to fermionic mass dimension one field to evade phenomenological physical bounds, the fermion-photon interaction carries a unique signature that may concretely appear at DM-searching running experiments    \cite{Lee:2015sqj,Alves:2014kta,Dias:2010aa,Agarwal:2014oaa,Duarte:2017svd}.  In particular, unitarity imposes subtle constraints on data from LHC, in analyzing dark matter scatterings in monophoton events \cite{Alves:2017joy,Alves:2014qua}. 
The coupling between fermionic mass dimension one field and the photon, to be experimentally yielded in monophoton events at the LHC, can explain the observed DM relic abundance. In this context, the XENON 1T experiment and its recently improved version XENON nT may probe nucleon-mass dimension one fermion scattering associated with virtual photon exchange. These DM search experiments have the goal of  detecting DM using chambers with a liquid xenon  target \cite{XENON:2019ykp}, with scintillation detectors \cite{Aprile:2022vux}.   There is a promising recent careful simulation research, regarding phonon-mediated measures \cite{Sassi:2022njl}.

Considering the fermion field, the ELKO \textcolor{black}{spinor expansion coefficients} implement the microscopic essence of DM as an elementary particle that has all the required properties to describe its defining features. The ELKO setup provides a dark sector in the extended Standard Model  \cite{daRocha:2011yr,daRocha:2009gb,deBrito:2019hih,Bonora:2014dfa,Pereira:2017efk,Rodrigues:2005yz}. Differently of Dirac and Majorana fermions, respectively regulated by the Dirac and Majorana first-order partial differential equations (PDEs),  ELKO are governed by a coupled system of first-order PDEs, mixing the four types of ELKO under the Dirac operator. Still, each type of ELKO does satisfy the Klein--Gordon equation. The four types of ELKO at rest are 
 constructed upon the fact that they are eigenspinors of the charge conjugation operator with dual helicity. Therefore, ELKO are split into two eigenclasses of self-conjugate and anti-self-conjugate spinors, under the charge conjugation operator. Both  subclasses are subsequently  split into two subclasses, involving the sign of the  right and left ELKO components under the action of the helicity operator \cite{Ahluwalia:2004ab,Ahluwalia:2004sz}. 
ELKO spinors naturally have spin sums completely different from the Dirac fermionic structure, presenting a more intricate form that takes into account an odd-parity operator, which also forces ELKO to have a generalized spinor conjugate that mixes the four types of ELKO, again differing from the Dirac spinor conjugate. These properties are indeed fundamental for the intrinsic darkness that is inherent to \textcolor{black}{ mass dimension one fermions}. This structure can circumvent a couple of issues on some of the existing models, which describe DM as composed of fermionic fields that satisfy some properties with no reference to a specific candidate. 
These types of models, describing light DM, are not able to occupy a state of  thermal
equilibrium with other particles along the emergence of the large-scale structure, as well as the nucleosynthesis era \cite{Pereira:2017efk}. These required properties can be implemented in a scenario where DM  fermionic fields, like the one associated with ELKO spinor, are endowed with a conserved dark charge. Hence, there are no renormalizable couplings among fermionic mass dimension one field and elementary particles in the Standard Model. Mass dimension one fermions support the DM halo surrounding galactic nuclei by a (quantum) degeneracy pressure effect mechanism. 
The dark matter effective mass between 0.1 KeV and 0.2 KeV can explain the mass-ratio relation for dwarf galaxies as well as the rotation curves of large galaxies can be modeled with the  fermion mass  $\sim$ 23 eV \cite{Ahluwalia:2022ttu,Pereira:2021dkn}. On the other hand, for the case of laboratory DM searches associated to sample nucleon recoils,  an excess in event rate for a  WIMP with a mass of $5$ GeV seems to be recently inferred in  simulations over a large quantity of experimental data  \cite{Sassi:2022njl}. Phonon-mediated detection was considered with a careful analysis of energy losses for such processes. Regarding the XENON 1T experiment, it  has a sensitivity to a wider range of WIMP masses, including lower ones, finding an excess associated with a possible DM  candidate with  $2.3$  KeV mass \cite{XENON:2020rca}. However, the new improved experiment XENON nT \cite{Aprile:2022vux} concluded that it was just a tritium background effect. \textcolor{black}{Taking into account the recent achievements associated with the Wigner degeneracy for mass dimension one fields \cite{Ahluwalia:2022yvk}, the previous phenomenological predictions should be updated. However, qualitative discussions can be maintained. Then, various new predictions associated with current experimental data  are open possibilities. }\\
\indent New developments on DM-photon tree-level scatterings were investigated in Ref. \cite{deGracia:2022enm}  in the context of QFTs with Maxwell and  Podolsky sectors \cite{Podolsky:1942zz,Podolsky:1944zz}. 
The Podolsky sector arises in the context of generalized quantum electrodynamics with additional gauge freedom coming from higher-order field equations, which permits removing eventual divergences as the one on computing the electron self-energy  \cite{Bertin:2009gs}. Introducing a Podolsky sector consistently preserves gauge invariance  \cite{Galvao:1986yq,Cuzinatto:2005zr} and the generalized quantum electrodynamics encompasses both gauge-invariant massive and massless photons  since it consists of the sole possibility of extending the standard (Maxwell) quantum electrodynamics to achieve a linear theory 
that is still covariant and invariant under Abelian gauge transformations \cite{Bufalo:2010sb,El-Bennich:2020aiq}. 
In the context of generalized quantum electrodynamics with a  Podolsky sector, an interacting theory with a positive total decay rate leads, for the massive sector, to the so-called Merlin modes \cite{Donoghue:2021eto}. Then, unitarity is preserved and just a small causality violation is introduced harmlessly since these exponentially decaying massive excitations can be interpreted as virtual particles. These internal lines improve the quantum properties of the model.

Our goal is to use the path integral formalism to evaluate  the vacuum-to-vacuum fermion-photon scattering  amplitude, also employing the Faddeev--Senjanovic Hamiltonian path-integral method \cite{Faddeev:1969su,Senjanovic:1976br}, consisting of a  generalization of the Feynman integral for any singular Lagrangian. In fact, a quantum field theory with constraints was investigated by Dirac himself \cite{Dirac:1958jc}, showing the first- and second-class constraints splitting of the algebra of Berezin brackets determines a division of constraints into two classes: the so-called first-class constraints and second-class ones. Constraints of first-class  present vanishing Berezin brackets, when computed with any other constraint in the  phase space. Otherwise, they represent  second-class constraints. The Faddeev--Senjanovic protocol quantizes field theories presenting first- and second-class constraints and will be here used to evaluate DM-photon scattering amplitudes in generalized quantum electrodynamics. The Hamiltonian analysis underlying fermionic mass dimension one field scattering and its functional quantization, to be also developed in this work, are completely original achievements in the literature. 
 
  One of the main aims of this work consists of engendering  a robust connection between complete formal quantum structures for models involving fermionic mass dimension one fields and the absence of observed signatures of light due to DM and photon scatterings, which are severely constrained in experiments \cite{Arina:2020mxo}. The mass dimension one fermionic   description of DM can be evinced,  since this setup is  intrinsically dark even if considering the full radiative corrections for the model. The definition of darkness comprises the feature that external light signs must be excluded. In this paper, our definition is sharper than this. Namely, we are going to consider the definition of darkness related to the vanishing of non-polarized Compton and pair annihilation amplitudes at 1-loop. This occurs in the case of a suitable set of phenomenological conditions. Despite these features,  processes with intermediate photon exchange are still allowed, justifying the current scrutiny here presented. It sheds new light on the possible detection of DM, via interaction with ordinary matter from laboratory devices. There is an increasing amount of achievements in this field  \cite{Sassi:2022njl,Agarwal:2014oaa}.

In this work, we investigate the Pauli-like interaction between \textcolor{black}{fermionic mass dimension one fields} and generalized photons in the Stueckelberg formalism, exploring for the first time in the literature general aspects of the DM-photon quantum interaction in a functional approach. First, to write the transition amplitude, it is  necessary to investigate carefully the constraint setup of the model. In the second level of analysis, we present not only the Schwinger-Dyson equations with the skeleton diagrams phenomenology of the electrodynamics investigated but also the Ward--Takahashi-like identities, preparing the way to investigate a mechanism for darkness. Finally, by the study of the vertex, Compton, and pair annihilation processes, with the renormalization procedure, it is possible to furnish a reliable background about the DM interpretation and find out a mechanism that avoids external light signals. This is based on vertex calculations developed up to $2$-loop order. This paper is organized as follows: 
Sec. \ref{sec1} revisits the ELKO setup,
fixing the standard notations and defining the ELKO spinors, its generalized dual, the spin sums, and the equations of motion governing ELKO, together with the twisted conjugation. The Feynman--Dyson propagator for  the \textcolor{black}{fermion mass dimension one field} is also deployed, and the Hamiltonian formulation is discussed. 
Sec. \ref{classical} introduces mass dimension one fields in the context of the Podolskyan intermediate bosons with Maxwell and Proca sectors, and the addition of a Stueckelberg scalar field. Ward--Takahashi-like  identities are analyzed from the Berezin brackets involving the physical fields in the theory and the first- and second-order Dirac constraints. 
The associated propagator is derived, containing a Merlin mode. Sec. \ref{quantum} is dedicated to presenting the analysis of the quantum transition amplitude, in the context of the phase space path integral for the vacuum-to-vacuum transition amplitude, within the Faddeev--Senjanovic method.
Due to the intricacies involving ELKO \textcolor{black}{spinor coefficients defining the field basis}, the 
results here derived are quite relevant and 
fruitful for the formulation of a mass dimension one fermionic description of DM. The  Dyson--Schwinger and the complete quantum equations as well as the Ward--Takahashi-like identities and the quantum gauge symmetry are scrutinized in the context of the present interaction. 
In Sec. \ref{compton}, the concept of  darkness is analyzed in the context of  scattering amplitudes   for Compton-like processes involving fermionic mass dimension one fields considering complete radiative 1-loop corrections. A useful quantitative analysis is also prescribed. Sec. \ref{secre} is devoted to the account of renormalizability and superrenormalizability
of the theory describing DM-generalized photon interactions in a Podolskyan scenario. Sec. \ref{conclu} devises the conclusions and achievements of this work, providing 
relevant perspectives, directions, and applications.

\section{ELKO spinor underlying framework and ramifications}
\label{sec1}
To approach QFT, in general, \textcolor{black}{after parity is incorporated}, one takes a spinor $\psi(p^\mu)$ in the $\left(\frac12,0\right)\oplus\left(0,\frac12\right)$ irreducible representation of the Lorentz group to constitute spin-1/2 quantum fields, which in the Weyl representation of the gamma matrices reads \cite{Ahluwalia:2022ttu}
	\begin{equation}	\psi(p^\mu) = \left( \begin{array}{c}
		\upphi_{\mathcal{R}}(p^\mu)\\
		\upphi_{\mathcal{L}}(p^\mu)
		\end{array}
		\right).
\end{equation}
The right and left Weyl spinors transform, respectively, by the $\left(\frac12,0\right)$ and  the $\left(0,\frac12\right)$ representations.\\
\indent The different chiral components ${\mathcal{R}}$ and ${\mathcal{L}}$ can be split off  by the chiral projectors\footnote{The standard notation is used, with $\gamma_5$ defined in terms of the gamma matrices in Weyl representation.}
\begin{equation} P_{{\mathcal{R}}/{\mathcal{L}}}\equiv \frac{1}{2}\big( \mathbb{I}\pm \gamma_5\big).    \end{equation}
 It is interesting  mentioning that the Dirac equation, in its massless limit, implies that $\upphi_{{\mathcal{R}}/{\mathcal{L}}}(p^\mu)=\upphi_\pm(p^\mu)$ with the latter denoting the helicity eigenstates, displayed as 
\begin{equation}
(\s\cdot\widehat \p) \upphi_\pm(p^\mu) = \pm \upphi_\pm(p^\mu).
\end{equation}
In spherical coordinates, 
\begin{align}
\upphi_+(k^\mu) & = \sqrt{m}  \left(
									\begin{array}{c}
									\cos\left(\frac{\theta}{2}\right)e^{- i \phi/2}\\
									\sin\left(\frac{\theta}{2}\right)e^{i \phi/2}
								\end{array}
	\right),\qquad\qquad
\upphi_-(k^\mu) = \sqrt{m}   \left(		\begin{array}{c}
									- \sin\left(\frac{\theta}{2}\right)e^{- i \phi/2}\\
									 \cos\left(\frac{\theta}{2}\right)e^{i \phi/2}
											\end{array}
									\right),
\end{align}
for the rest frame $k^\mu=\lim_{\vec p\to0}p^\mu$ \cite{Ahluwalia:2022ttu}.

ELKO are  eigenspinors of the charge conjugation operator, with their chiral components presenting reverse helicity.  For $\Uptheta   =-i\sigma_2$ denoting the Wigner time-reversal operator, the construction of ELKO regards a direct sum of a right-handed Weyl spinor $\upphi_\pm$ with fixed helicity with a left-handed  spinor $\Uptheta \upphi_\pm^\ast$, with opposite helicity
\begin{equation}
 \s\cdot\widehat{\p} \left[\Uptheta \upphi_\pm^\ast(p^\mu)\right]  = \mp
  \left[\Uptheta \upphi_\pm^\ast(p^\mu)\right]  .
 \end{equation}
\indent Expressing the charge conjugation involutive operator as  \cite{Ahluwalia:2019etz}
\begin{equation}
	\mathsf{C} = \left(\begin{array}{cc}
					\0_{2\times 2} & i\Uptheta \\
					-i \Uptheta & \0_{2\times 2}
					\end{array}
					\right) \mathsf{K}, 
	\end{equation}
where the $\mathsf{K}$ operator implements complex conjugation, the self-conjugate ($\uplambda^{{\scalebox{0.65}{$S$}}}(p^\mu)$)  and anti-self-conjugate ($\uplambda^{{\scalebox{0.65}{$A$}}}(p^\mu)$), the ELKO spinor respectively  satisfy, 
\begin{equation}
\mathsf{C} \uplambda^{S}_\pm(p^\mu) =  +  \uplambda^{S}_\pm(p^\mu),\qquad
\mathsf{C} \uplambda^{A}_\pm(p^\mu) =   - \uplambda^{A}_\pm(p^\mu).
\end{equation}
Consequently, the ELKO at rest can be expressed as $\lim_{{{\scalebox{0.75}{$\p$}}}\to 0}\uplambda_\upalpha(p^\mu)$, as \cite{Ahluwalia:2019etz}
\begin{align}
 &\uplambda^{{\scalebox{0.65}{$S$}}}_\pm  =  \lim_{{{\scalebox{0.75}{$\p$}}}\to 0}\left(
					\begin{array}{c}
					+i \Uptheta\left[\upphi_\pm(p^\mu)\right]^\ast\\
								\upphi_\pm(p^\mu)
					\end{array}
					\right),  \qquad\qquad\uplambda^{{\scalebox{0.65}{$A$}}}_\pm  =  \lim_{{{\scalebox{0.75}{$\p$}}}\to 0}\left(
					\begin{array}{c}
					- i \Uptheta\left[\upphi_\mp(p^\mu)\right]^\ast\\
								\upphi_\mp(p^\mu)
					\end{array}
					\right).\label{sa}
				\end{align}
				In this sense, one can consider the ELKO spinor space  as the direct sum $S = S_{{{\scalebox{0.6}{\textsc{S}}}}} \oplus S_{{{\scalebox{0.6}{\textsc{A}}}}}$
				of 
two vector subspaces of self-conjugate and anti-self-conjugate ELKO, each one consisting of an invariant spinor eigenspace under the action of the charge conjugation operator.   Defining parity as  $\mathsf{P}= m^{-1} \gamma^\mu p_\mu$  \cite{Speranca:2013hqa} and the time-reversal operator $\mathsf{T}= -i \gamma_{5}\mathsf{C}$ yields $
 (\mathsf{C}\mathsf{P}\mathsf{T})^2 = \I$, and $\left \{\mathsf{C},\mathsf{P}\right\}=0$. This algebra of operators acting on ELKO allocates the ELKO as non-standard spinors in the Wigner and Lounesto classifications \cite{daRocha:2008we,Ablamowicz:2014rpa,daRocha:2007sd,Fabbri:2016msm}.\\
\indent The four types of ELKO are well known not to satisfy Dirac equations, but the following coupled system of first-order field equations \cite{Ahluwalia:2019etz}, 
\begin{align}
&\gamma_\mu p^\mu \uplambda^{{\scalebox{0.65}{$S$}}}_\pm(p^\mu) = \pm im \uplambda^{{\scalebox{0.65}{$S$}}}_\mp(p^\mu),\qquad\quad
 \gamma_\mu p^\mu \uplambda^{{\scalebox{0.65}{$A$}}}_\pm(p^\mu) =\mp  im \uplambda^{{\scalebox{0.65}{$A$}}}_\mp(p^\mu),\label{coupled}
\end{align}
which can be straightforwardly obtained when one acts the Dirac operator $\gamma^\mu p_\mu$ on the four types of ELKO in Eqs. (\ref{sa}), which satisfy the Klein--Gordon equations \cite{Ahluwalia:2004sz}.\\
 \indent ELKO spinors for arbitrary momentum are then encoded by taking into account a boost 
 $
\uplambda^{S/A}_{\pm}(p^\mu) = \mathsf{D}(L(p^\mu)) \; \lim_{{{\scalebox{0.75}{$ \p$}}}\to 0}\uplambda^{S/A}_{\pm}(p^\mu),$ where $\mathsf{D}(L(p^\mu))$ is a representation of the Lorentz group element $L(p^\mu)$ at arbitrary momentum. Denoting by $a(\p)$ and $b(\p)$ annihilation operators, and $a^\dagger(\p)$ and $b^\dagger(\p)$  creation operators, 
the free \textcolor{black}{ fermionic mass dimension one} quantum field is given by  \cite{Ahluwalia:2022ttu,Ahluwalia:2019etz,Ahluwalia:2008xi}, 
\begin{equation}
\mathfrak{f}(x) =\int
\frac{d^3 p}{(2\pi)^3}
\frac{1}{({2 m E(\p)})^{1/2}}
\sum_\upalpha
\left[
a_\upalpha(\p)
\varepsilon_\upalpha(\p) e^{-i p^\mu x_\mu}
+
b^\dagger_\upalpha(\p)
\chi_\upalpha(\p) e^{i p^\mu x_\mu}
\right].                   \label{eq113}
\end{equation}
in which the expansion coefficient basis $\varepsilon_\upalpha(\p)$ and $\chi_\upalpha(\p)$, associated to ELKO spinors, will be properly defined latter.\\
\indent The annihilation and creation operators satisfy the non-vanishing canonical anti-commutators
\begin{equation}
\left\{a_{\upalpha}(\p),a^{\dag}_{\upalpha'}(\p')\right\}=\left\{b_{\upalpha}(\p),b^{\dag}_{\upalpha'}(\p')\right\}=(2\pi)^{3}\delta_{\upalpha\upalpha'}\delta^{3}(\p-\p'),
\end{equation}
 The spinor field $\mathfrak{f}(x)$ has {mass dimension} 
one and fermionic statistics. Hence, \textcolor{black}{mass dimension one} spin $1/2$ fields cannot partake in Standard Model doublets, therefore constituting first-principle candidates to describe DM \cite{Ahluwalia:2022ttu,Ahluwalia:2019etz}. \textcolor{black}{Although  our focus is on}  \textcolor{black}{fermionic dark matter},  \textcolor{black}{the structure developed in this paper can be immediately generalized for comprising a wide set of spin $1/2$ mass dimension one fields, including also the bosonic ones \cite{Ahluwalia:2022ttu}}. \textcolor{black}{Interestingly, the associated expansion coefficients also depend on the ELKO spinor structures grouped in a different basis}.\\
\indent Regarding the fermion field, the self-conjugate sector has the following basis \cite{Ahluwalia:2022yvk}
\bea        \varepsilon_1(p)=\uplambda^{S}_{+}(p)  \ ,\quad \ \varepsilon_2(p)=\uplambda^{S}_{-}(p) \ ,\quad \ \varepsilon_3(p)=-i\uplambda^{A}_{+}(p) \ ,\quad \ \varepsilon_4(p)=-i\uplambda^{A}_{-}(p),  \label{basis}     \eea
and the anti-self-conjugate one has
     \bea        \chi_1(p)=\uplambda^{A}_{+}(p)  \ ,\quad \  \chi_2(p)=\uplambda^{A}_{-}(p) \ ,\quad \ \chi_3(p)=-i\uplambda^{S}_{+}(p) \ ,\quad \  \chi_4(p)=-i\uplambda^{S}_{-}(p).       \eea
     with both basis being defined in terms of the ELKO spinors.\\
\indent According to Ref. \cite{Wigner:1939cj}, it is possible to assume a two-fold degeneracy in the presence of an anti-linear operator, doubling the degrees of freedom. This is an irreducible representation, since the rotational constraint demanded for one particle states can only be fulfilled by such structure. The non-covariant parts are explicitly cancelled in the spinor products. As a consequence, the associated spin sums are covariant without the need for the so-called $\tau$-deformation  \cite{Ahluwalia:2022ttu}. Additionally, it is worth mentioning that considering the new basis and the behavior under parity operation, the Lee-Wick no-go theorem can be evaded on the same grounds as in Ref. \cite{HoffdaSilva:2019eao}.\\ 
        \indent Since the norm of ELKO spinors using the Dirac dual equals zero, a new definition for the dual  is necessary. It reads
        \bea   \gdualn \varepsilon_\sigma(p)=\Big[{\mathcal{P}}\varepsilon_\sigma    \Big]^\dagger(p)\gamma_0, \quad \quad \gdualn \chi_\sigma(p)=\Big[-{\mathcal{P}}\chi_\sigma    \Big]^\dagger(p) \gamma_0,    \eea 
with ${\mathcal{P}}$ representing the parity operator \cite{Speranca:2013hqa}. The non-vanishing products are given by 
\begin{align}
& \gdualn \varepsilon_\upalpha(p^\mu) \varepsilon_{\upalpha^\prime}(p^\mu)
 =  m \delta_{\upalpha\upalpha^\prime}= - \gdualn \chi_\upalpha(p^\mu) \chi_{\upalpha^\prime}(p^\mu)
 \label{eq117}
\end{align}
with spin sums 
\begin{align}
& \sum_\upalpha \varepsilon_\upalpha(p^\mu) \gdualn{\varepsilon}_\upalpha(p^\mu) =  m \I = - \sum_\upalpha \chi_\upalpha(p^\mu) \gdualn{\chi}_\upalpha(p^\mu)\label{sum},
\end{align}
and completeness relation $
\sum_{\upalpha}\left[ \varepsilon_\upalpha(p^\mu) \gdualn{\varepsilon}_\upalpha(p^\mu) - \chi_\upalpha(p^\mu) \gdualn{\chi}_\upalpha(p^\mu)\right]  = 2m\I.$
To prospect the causal structure satisfied by the \textcolor{black}{mass dimension one} field $\mathfrak{f}(x)$  in Eq. (\ref{eq113}), its adjoint is naturally defined as
\begin{equation}
{\gdualn{\mathfrak{f}}}(x) =
\int
\frac{d^3 p}{(2\pi)^3}
\frac{1}{\sqrt{2 m E(\p)}}
\sum_\upalpha\left[
a^\dagger_\upalpha(\p) \gdualn{\varepsilon}_\upalpha(\p) e^{i p\cdot x}
+ b_\upalpha(\p) \gdualn{\chi}_\upalpha(\p) e^{- i p\cdot x}
\right],\label{dual}
\end{equation}
with   
$
\{ {\gdualn{\mathfrak{f}}}(x) , {\mathfrak{f}}(x^\prime)\}=0$, when $x$ and $x'$ are separated by space-like intervals. This result is in compliance with the
fermionic statistics for the creation and annihilation operators, yielding a consistent local QFT. 
Ref. \cite{Ahluwalia:2019etz} showed that the Feynman--Dyson propagator for mass dimension one fermionic fields reads
\begin{align}
S_{{\scalebox{0.6}{$\textsc{FD}$}}}(x^\prime-x)  =
\int\frac{d^4p}{(2\pi)^4}\frac{\mathsf{I}_{4\times 4}}{\Box + m^2 + i\xi}e^{i p\cdot(x - x^\prime)},
\label{eqp}\end{align} where $\mathsf{I}_{4\times 4}$ denotes the identity operator in  spinor space, and the free field Lagrangian density is given by
\begin{equation}
\mathcal{L}_{{\scalebox{0.58}{$\textsc{free}$}}}(x)  =\partial_\mu {\gdualn{\mathfrak{f}}} (x)
\partial^\mu\mathfrak{f}(x)
 - m^2 {\gdualn{\mathfrak{f}}}(x) \mathfrak{f}(x). \label{eq:Lagrangian}
\end{equation} 
In QFT,  transition probabilities can be computed by means of the usual Hermitian conjugation. It is associated to the obtainment of amplitudes and correlated observables. 
However, the specific dual ELKO \textcolor{black}{spinor} structure demands a generalized Hermitian prescription. Therefore, Ref. \cite{Ahluwalia:2022ttu} introduced the involutive twisted conjugation $\ddag$  to calculate  transition probabilities and observables 
\begin{equation}
\left[\uplambda^{S/A}_{\upalpha}(\p)\right]^{\ddag}=-i\upalpha\left[\uplambda^{S/A}_{-\upalpha}(\p)\right]^{\dag},\label{eq119}
\end{equation}

\indent The Lagrangian is invariant by the generalized twisted conjugation $\mathcal{L}_{{\scalebox{0.58}{$\textsc{free}$}}}^{\ddag}(x)=\mathcal{L}_{{\scalebox{0.58}{$\textsc{free}$}}}(x)$ leading to a positive definite Hamiltonian with the standard form. For mass dimension one fermions, the usual demand for hermiticity must be then  replaced by requiring  invariance under the twisted conjugation $\ddag$. When acting on spinors different than ELKO, the twisted  conjugation is equivalent to the Hermitian one. Then, it is possible to discuss the optical theorem and the preservation of products taken with this conjugation \cite{Ahluwalia:2022ttu}. Most importantly, there is also a well-defined prescription to calculate positive probabilities from amplitudes obtained by the twisted conjugation procedure.

\indent Considering the Lagrangian structure, the conjugate momentum to $\mathfrak{f}(x)$ reads \cite{Ahluwalia:2022ttu,Ahluwalia:2019etz}
\begin{equation}
\uppi(x) =
\frac{\partial \mathcal{L}_{{\scalebox{0.58}{$\textsc{free}$}}}(x)}
{\partial       \dot{\gdualn{\mathfrak{f}}}   (x)   } =  
\dot{\mathfrak{f}}(x),
\qquad  \qquad \gdualn{\uppi}(x) =
\frac{\partial \mathcal{L}_{{\scalebox{0.58}{$\textsc{free}$}}}(x)}
{\partial       \dot{\mathfrak{f}}  (x)   } = 
\dot{\gdualn{{\mathfrak{f}}}}(x).
\end{equation}
Fermionic field locality is governed by the canonical  anti-commutators 
\begin{align}
\left\{
\mathfrak{f}(t,\vec x),\gdualn{\uppi}(t,\vec x^\prime)\right\} &=
i \mathsf{I}_{4\times 4}\delta^3(\vec x-\vec x^\prime), \qquad \left\{
\gdualn{\mathfrak{f}}(t,\vec x),\uppi(t,\vec x^\prime)\right\} = 
i\mathsf{I}_{4\times 4} \delta^3(\vec x-\vec x^\prime).
\end{align}

\indent   Considering  $\mathfrak{f}$ and $\gdualn{\mathfrak{f}}$ independent variables, similar to the analysis of the Dirac fermion Lagrangian case, the \textcolor{black}{mass dimension one fermionic field } admits the  Hamiltonian
\bea H=\int \frac{d^3p}{(2\pi)^3}E(\p)\Bigg[\sum_{\upalpha}\left(a^\dagger_\upalpha a_\upalpha+b^\dagger_\upalpha b_\upalpha\right)-2\Bigg],        \eea
which is analogous to the mass dimension three-halves fermionic case. \textcolor{black}{It is worth mentioning the fact that the zero-point energy of the fermionic mass dimension one field is exactly the opposite of the one associated with the so-called spin $1/2$ bosons with the same mass \cite{Ahluwalia:2022ttu}.  } The free fermion Lagrangian density admits a global U(1) symmetry. The corresponding fermion number  conserved charge
$N_{\mathfrak{f}} \sim  \gdualn{\mathfrak{f}}{\uppi} - \gdualn\uppi \mathfrak{f}$ follows from Noether's theorem, which can be interpreted as a dark charge associated with the mass dimension one fermion field with ELKO spinor expansion coefficients \cite{Pereira:2021dkn}.

\section{generalized photons and fermionic mass dimension one fields, in the first- and second-class constraints setup}

\label{classical}

Maxwell's classical electrodynamics is well known to regard first-order derivatives in the Lagrangian. The generalized electrodynamics proposed by Podolsky takes into account also the second-order  derivatives, being consistent with linearity and the standard Lorentz and U(1) symmetries of the theory \cite{Podolsky:1945chv}. 
Denoting the electromagnetic field strength by $F_{\mu\nu} = \partial_{\mu}A_{\nu}-\partial_\nu A_\mu$, where $A_\mu$ denotes the electromagnetic potential, the standard Podolsky gauge field  interacting with the mass dimension one field can be encoded into the Lagrangian\footnote{Throughout this work Greek indexes denote spacetime coordinates  $\mu=0,\ldots, 3$ and Latin indexes denote the spatial ones $i=1,2,3$.}
\beq
    {\cal{L}}_{{\scalebox{0.6}{$\textsc{pod}$}}}=-\frac{1}{4}F_{\mu \nu}F^{\mu \nu}-\frac{a^2}{2}B_\mu B^\mu+a^2\partial_\mu B_\nu F^{\mu \nu}+\partial_\mu {\gdualn{\mathfrak{f}}} \partial^\mu {\mathfrak{f}}-m^2\gdualn{\mathfrak{f}} {\mathfrak{f}}+\frac{ig}{2}{\gdualn{\mathfrak{f}}} \sigma_{\mu \nu} {\mathfrak{f}} F^{\mu \nu},\label{lpod1}
\eeq
with $a^2=\frac{1}{M^2}$,  for $M$ denoting the Podolsky photon mass,  $B_\mu$ denoting an auxiliary field\footnote{Completing the square in the $B_\mu(x)$ field and eliminating a non-dynamical field, one obtains the standard form of the Podolsky Lagrangian.} and the definition $\sigma_{\mu \nu}=\frac{1}{2}\left[\gamma_\mu,\gamma_\nu \right]$ was used. For achieving compatibility between the theory and experiments, the following lower bound for the mass of the Podolsky mass $M\gtrsim 370$ GeV can be considered, due to constraints from Bhabha scattering regarding its role on generalized QED$_4$  phenomenology \cite{Bufalo:2014jra}. The effective Bhabha scattering in the Podolsky framework is similar to considering a   Feynman regulator cutoff parameter, which can be identified as the Podolsky mass. Therefore, the  electron-positron scattering with 12 GeV $\lesssim$ $\sqrt{s}\lesssim  46.8$ GeV yields the bound $M\gtrsim 370$ GeV. Important techniques involving  generating functionals and partition functions demonstrated that the  self-energy of the electron and the vertex sector of the generalized quantum electrodynamics are both finite at the ultraviolet regime, also complying with renormalizability \cite{Bufalo:2010sb}.  The Lagrangian (\ref{lpod1}) is invariant under the twisted conjugation. 
One can redefine the electromagnetic potential $A_\mu \mapsto A_\mu+a^2 B_\mu$ and, together with the mapping $a^2 B_\mu \mapsto B_\mu$, the Lagrangian (\ref{lpod1}) can be rewritten, using the notation $\theta_{\mu \nu}(p)=g_{\mu \nu}-\frac{\partial_\mu \partial_\nu}{\Box}$, as 
\beq
   \!\!\!\!\!\! {\cal{L}}_{{\scalebox{0.6}{$\textsc{pod}$}}}\!&=&\!-\frac{1}{4}F_{\mu \nu}F^{\mu \nu}-\frac{1}{2} \left[M^2\left(B_\mu+\frac{1}{M}\partial_\mu \phi\right)^2+B^\mu\Box \theta_{\mu \nu}B^\nu \right] +\partial_\mu {\gdualn{\mathfrak{f}}} \partial^\mu {\mathfrak{f}}-m^2{\gdualn{\mathfrak{f}}} {\mathfrak{f}}\nonumber\\&&+\frac{ig}{2}{\gdualn{\mathfrak{f}}} \sigma_{\mu \nu} {\mathfrak{f}} \big(F^{\mu \nu}+G^{\mu \nu}\big)={\cal{L}}_{{\scalebox{0.6}{$\textsc{pod}$}}}^\ddagger,
\label{pod2}\eeq
which is invariant  under the twisted conjugation,
where $G_{\mu\nu} = \partial_{\mu}B_{\nu}-\partial_\nu B_\mu$. A non-observable Stueckelberg scalar particle has been introduced, for the system to generate an extra gauge symmetry to describe in a covariant way the three possible polarizations of the massive vector field. The Stueckelberg field $\phi$  reinstates the gauge symmetry 
broken by the addition of a massive term \cite{Ruegg:2003ps}. It consists of a useful trick to derive Ward--Takahashi-like identities. The local two independent local symmetries are represented by the invariance under the transformations
\bea \delta A_\mu(x) =\partial_\mu \alpha(x), \qquad \quad  \delta B_\mu(x) =\partial_\mu \Lambda(x), \qquad \quad\delta \phi(x)=-M\Lambda(x).       \eea 
Therefore, one ends up with a theory exhibiting a {U}(1)$\times$ {U}(1) local symmetry. 
The non-vanishing Berezin brackets read
\begin{subequations}
\beq
\!\!\!\!\!\!\!\!\left\{
\mathfrak{f}(t,\vec x),\gdualn{\uppi}(t,\vec x^\prime)\right\}_B &=&-
 \mathsf{I}_{4\times 4}\delta^3(\vec x-\vec x^\prime), \qquad \left\{
\gdualn{\mathfrak{f}}(t,\vec x),\uppi(t,\vec x^\prime)\right\}_B = -
\mathsf{I}_{4\times 4} \delta^3(\vec x-\vec x^\prime),  \\
\left\{
A_\mu(t,\vec x),\Pi^{\nu  A}(t,\vec x^\prime)\right\}_B &=&
 \delta_\mu^{\ \nu}\delta^3(\vec x-\vec x^\prime), \qquad \left\{
B_\mu(t,\vec x),\Pi^{\nu B}(t,\vec x^\prime)\right\}_B = 
\delta_\mu^{\nu} \delta^3(\vec x-\vec x^\prime),
\eeq
\end{subequations}
whereas for the Stueckelberg scalar field, it yields
\bea \left\{
\phi(t,\vec x),p_\phi(t,\vec x^\prime)\right\}_B=\delta^3(\vec x-\vec x^\prime).\eea
\noindent The Berezin brackets are defined as 
\bea  \{F,G\}_B=\int d^4z\left[\frac{\partial_r F}{\partial \Phi_A(z)}\frac{\partial_l G}{\partial \mathfrak{p}_A(z)}-(-1)^{P_FP_G}\frac{\partial_r G}{\partial \Phi_A(z)}\frac{\partial_l F}{\partial \mathfrak{p}_A(z)}\right] \eea
in terms of right- and left-  derivatives in which the $P$ denotes the Grassmann parity of the field valued function, $r$ and $l$ denote the right and left  derivatives, respectively. The $\mathfrak{p}_A(x)$ and $\Phi_A(x)$ are collective designations for all the momenta and the fields.

\indent The canonical momenta are given by
\begin{subequations}
\beq \Pi^{i A}(x)&=&F_{0i}(x)-ig{\gdualn{\mathfrak{f}}}(x) \sigma_{0i}{\mathfrak{f}}(x),\qquad  \quad \Pi^{0A}(x)=0,\\ \Pi^{i B}(x)&=& -G_{0i}(x)-ig{\gdualn{\mathfrak{f}}}(x) \sigma_{0i}{\mathfrak{f}}(x),\qquad  \quad \Pi^{0B}(x)=0.
\eeq
\end{subequations}
The momenta for the fermionic sector have been defined in Sec. \ref{sec1}, whereas  the definition for the  vector bosonic sector reads  \beq\mathfrak{p}^A(x)=\frac{\partial {\cal{L}}_{{\scalebox{0.6}{$\textsc{pod}$}}}}{\partial \dot \Phi_A(x)},\eeq with $\Phi_A(x)$ being a collective designation for the bosonic fields. For the fermionic mass dimension one quantum field, it reads
\beq \gdualn{\uppi}(x)= \dot{ {\gdualn{\mathfrak{f}}}}(x),\qquad  \quad\uppi(x)=\dot{\mathfrak{f}}(x), \eeq
whereas for the bosonic scalar sector, it is given by  
\bea p_\phi(x)=\dot{\phi}(x)+MB_0(x). \eea

 The canonical Hamiltonian density can be written explicitly as
\beq
\mathcal{H}_C&=&p_\phi \dot \phi+\Pi^{iA} \dot A_i+\Pi^{iB}\dot B_i+\dot{\gdualn{\mathfrak{f}}} \uppi+\gdualn{\uppi} \dot {\mathfrak{f}} -\mathcal{L}_{{\scalebox{0.6}{$\textsc{pod}$}}} \nonumber \\
&=&\gdualn{\uppi} \uppi+\frac{1}{2}p^2_\phi-MB_0p_\phi+   \nabla {\gdualn{\mathfrak{f}}}\nabla {\mathfrak{f}} +m^2{\gdualn{\mathfrak{f}}} {\mathfrak{f}} +\frac{1}{2}\Pi^A_i\Pi^A_i-\frac{1}{2}\Pi^B_i\Pi^B_i+\Pi^{i  A}\partial_iA_0+\Pi^{iB}\partial_iB_0\nonumber \\
&&+\frac{1}{4}F_{ij}F_{ij}\!-\!\frac{1}{4}G_{ij}G_{ij}\!+\!ig\big(\Pi^{iA}\!-\!\Pi^{iB}\big) {\gdualn{\mathfrak{f}}} \sigma_{0i}{\mathfrak{f}}\!-\!\frac{ig}{2}{\gdualn{\mathfrak{f}}} \sigma_{ij} {\mathfrak{f}} \big(F_{ij}\!+\!G_{ij}\big)\!-\!\frac{M^2}{2}\!\left(\!B_i\!+\!\frac{\partial_i \phi}{M}\right)^2, \eeq
corresponding to the following primary constraints 
\begin{align}
\Sigma_1\left(x\right)\equiv&\,\Pi_0^A\left(x\right)\approx 0,\qquad\qquad\qquad
\Sigma_2\left(x\right)\equiv \Pi_0^B\left(x\right)\approx 0,
\end{align}
whose time evolution implies new constraints. 
\beq \Sigma_3\equiv \dot \Sigma_1&=& \left\{\Sigma_{1}(x),\int \mathcal{H}_P(y)d^3y\right\}_P=\partial_i \Pi_i^{A}(x)\approx 0,\label{sig3}\\
 \Sigma_4\equiv \dot \Sigma_2 &=& \left\{\Sigma_{2}(x),\int \mathcal{H}_P(y)d^3y\right\}_P=\partial_i \Pi_i^{B}(x)+Mp_\phi(x)\approx 0,\label{sig4} \eeq
with  $\mathcal{H}_P(y)$ denoting the primary Hamiltonian defined as 
\bea \mathcal{H}_P(y)=\mathcal{H}_C(y)+\Lambda_1(x)\Sigma_1(x)+\Lambda_2(x)\Sigma_2(x),       \eea
for $\Lambda_1(x),\Lambda_2(x)$ denoting Lagrange multipliers.

 The time evolution of these secondary constraints does not lead to new constraints and the Dirac--Bergmann algorithm, converting the singular
Lagrangian  into a constrained Hamiltonian comes to its end. The secondary Hamiltonian,
\bea \mathcal{H}_S(y)=\mathcal{H}_P(y)+\Lambda_3(x)\Sigma_3(x)+\Lambda_4(x)\Sigma_4(x),        \eea
in which $\Lambda_3(x),\Lambda_4(x)  $ being Lagrange multipliers, was considered. None of the $\Lambda_a(x)$, for $a=1,\ldots,4$, had to be  determined in this process, meaning a first-class theory.  Therefore, to obtain a uniquely defined system, extra constraints must be considered, the gauge fixing ones to turn the gauge field sector into a second-class one, eliminating all the local redundancy associated with canonical transformations,
\begin{subequations}
\begin{eqnarray}
 \Sigma_5(x)&\equiv& A_0-ig\frac{ \partial_i\big(\gdualn{\mathfrak{f}}\sigma_{0i}\mathfrak{f} \big)}{\nabla^2} \approx 0,\\\Sigma_6(x) &\equiv&  \partial_iA_i \approx 0, \\
 \Sigma_7(x)&\equiv& B_0-\frac{Mp_\phi-ig \partial_i\big(\gdualn{\mathfrak{f}}\sigma_{0i}\mathfrak{f} \big)}{\nabla^2-M^2} \approx 0,\\\Sigma_8(x) &\equiv&\partial_iB_i \approx 0.
\end{eqnarray}
\end{subequations}
It is interesting to mention that the condition on the divergence of the fields is first fixed. Then, the equations of motion imply the conditions for the zeroth coordinate of the bosonic fields. The inverse of differential operators is a formal notation to be understood in the Green function sense. 

 The time evolution of the constraints determines the entire set of Lagrange multipliers, eliminating the previous freedom associated with the gauge symmetry of the first-class system.  Therefore, a set of second-class constraints has been achieved, given by  $\omega_k=\left\{\Sigma_1,\Sigma_5,\Sigma_3,\Sigma_6,\Sigma_4, \Sigma_8,\Sigma_2,\Sigma_7\right\}$, as evidenced by the nonvanishing determinant of the matrix $\left[\Omega_{jk}\right]$, whose $jk^{\rm th}$ component reads  $\Omega_{jk}\equiv\left\{\omega_j\left(\vec{x},t\right),\omega_k\left(\vec{y},t\right)\right\}_P$, given by the ordered block-diagonal matrix,
\beq
-\sigma_2\oplus\left(\sigma_3\nabla^2\right)\oplus\left(\sigma_3\nabla^2\right)\oplus\sigma_2,
\eeq
written in terms of Pauli matrices, explicitly given by  
\begin{eqnarray}
  \left[\Omega_{jk}\right] &=& \left[\begin{array}{cccccccc}
                                 0 & -1 & 0 & 0 & 0 & 0 &0 &0 \\
                                 1 & 0 & 0 & 0 & 0 & 0 &0 &0  \\
                                 0 & 0 & 0 & \nabla^2 & 0 & 0 &0 &0  \\
                                 0 & 0 & -\nabla^2 & 0 & 0 & 0 &0 &0  \\
                                 0 & 0 & 0 & 0 & 0 & \nabla^2 &0 &0 \\
                                 0 & 0 & 0 & 0 & -\nabla^2 & 0 &0 &0 \\
                                 0 & 0 & 0 & 0 & 0 &0 &0 &-1 \\
                                 0 & 0 & 0 & 0 & 0 & 0 &1 &0 
                               \end{array}\right]\delta^{\left(3\right)}\left(\vec{x}-\vec{y}\right),
\end{eqnarray}
whose determinant is   $\det[\Omega_{jk}]=\det \big[\big(\nabla^2\big)^4\delta^{\left(3\right)}\left(\vec{x}-\vec{y}\right)\big]$.  Regarding the counting of the degrees of freedom, the gauge fixed model has eight constraints and a total of eighteen local excitations in the bosonic sector of the phase space. The fermionic one has no constraints whatsoever, which differs from the Dirac particle case, whose Lagrangian is constrained. In this latter case, the Lagrangian dictates its on-shell character and inner structure as well. \textcolor{black}{ The unconstrained fermion sector has sixteen phase space degrees of freedom, considering also the momentum variables. Therefore, the configuration space must be associated with eight degrees of freedom. This is in accordance with the dimensionality of the basis \eqref{basis} in compliance with Wigner degeneracy.}
These degrees of freedom are the external particles that can contribute to scattering processes for nucleon recoil experiments, whose aim is to detect DM.\\
\indent Therefore, since the \textcolor{black}{mass dimension one fermionic field} internal structure is constructed a priori, the Lagrangian just expresses that it obeys the Klein--Gordon equation. Considering these observations, this absence of fermionic constraints is an expected result. \textcolor{black}{Regarding the spinor coefficients}, a mixture of conjugated and self-conjugated ELKO  just obeys the Klein-Gordon and no other lower derivative equation, but a coupled system of first-order PDEs involving the self-conjugate and anti-self-conjugate ELKO \cite{Ahluwalia:2022ttu,Ahluwalia:2022yvk}. \textcolor{black}{One can easily note that the two sectors associated with the new basis \eqref{basis} are indeed the elements with opposite charge conjugacy}.   Therefore,  the fermionic configuration space is spanned by the following superposition of spinor expansion coefficients
\bea \sum_{\substack{\upalpha=1}}^4\mathfrak{C}_\upalpha^\varepsilon(p^\mu)\varepsilon_\upalpha(p^\mu) +\sum_{\substack{\upalpha=1}}^4\mathfrak{C}_\upalpha^\chi(p^\mu)\chi_{\upalpha}(p^\mu)\eea
\textcolor{black}{in which the sum is taken over the new basis elements associated with Wigner degeneracy. In accordance with this anti-linear setting, the coefficients $\mathfrak{C}_\upalpha^{ \ \varepsilon, \chi}(p^\mu)$ must be real functions. } \\
\indent Summing up, after all, there are five degrees of freedom, in configuration space, for the bosonic sector, being two for the massless photon and three for the massive one. The fermionic configuration space  has four degrees of freedom, the same as in the Dirac fermionic case. This indicates that, in terms of degrees of freedom, the scalar particle is surpassed by the massive gauge field and it does not contribute as a new local excitation. This is expected since it is a pure gauge field.

 It is interesting to note that the massive bosonic field seems to turn the Hamiltonian into a non-positive definite one. This is due to the well-known negative norm states of the free  Podolsky model \cite{Podolsky:1942zz,Bertin:2009gs}. On the other hand, if the theory is interacting, the complete physical part of the inverse propagator associated with $B_\mu(x)$, reads 
\bea     {\cal{P}}_{\mu \nu}^{-1}=-i\theta_{\mu \nu}(p^2-M^2_{{\scalebox{0.64}{$\textit{R}$}}})-i\mathcal{\pi}_{\mu \nu {{\scalebox{0.64}{$\textit{R}$}}}}^\intercal(p^2),\                                                  \eea
with the total contribution to the polarization tensor $\mathcal{\pi}_{\mu \nu {{\scalebox{0.64}{$\textit{R}$}}}}^\intercal(p^2)$ denotes the sum of the contributions of generalized QED$_4$ and an eventual one 
from the coupling with the mass dimension one field. The $M_{{\scalebox{0.64}{$\textit{R}$}}}$ denotes the renormalized Podolsky mass\footnote{{Hereon the   subindex $R$, also alternatively appearing as a superindex, will denote any renormalized scalar or field.}}.
It yields, up to longitudinal non-physical terms, 
\beq
{\cal{P}}_{\mu\nu}\sim \frac{i\theta_{\mu \nu}}{p^2-M^2_{{\scalebox{0.6}{$\textsc{P}$}}}-i|\upgamma|},              \eeq
considering that the total contribution to the polarization tensor is transverse 
$\pi_{\mu \nu {{\scalebox{0.64}{$\textit{R}$}}}}^\intercal(k)=\theta_{\mu \nu}\Pi^\intercal_{{\scalebox{0.64}{$\textit{R}$}}}(k)$, as it is going to be  proved further. 
The $|\upgamma|$ denotes the modulus of the 
imaginary part of $\Pi^\intercal_{{\scalebox{0.64}{$\textit{R}$}}}(k)$,  
evaluated at the pole mass $M_{{\scalebox{0.6}{$\textsc{P}$}}}$.
If this mass is such that this object is non-vanishing, 
the massive excitations present a Merlin mode behavior \cite{Donoghue:2021eto}. 
It does not violate unitarity since it becomes a special 
kind of backward traveling resonance and its internal
lines are never cut. This occurs since  an
exponentially decaying mode sets in, and it does not contribute to
asymptotic excitations. 
Therefore, although Merlin modes contribute to the renormalization of its parameters, it does not contribute
directly
to the asymptotic Hamiltonian operator, which has positive eigenvalues. Regarding this last statement, although the fermionic sector presents non-trivial features due to its twisted dual structure, its asymptotic free Hamiltonian operator has indeed a positive definite spectrum, according to the introductory remarks.
As it will be discussed, the renormalized photon self-energy is defined to generate a unit residue for the massless field. It implies that $M_{{\scalebox{0.64}{$\textit{R}$}}}$ is defined as $M_{{\scalebox{0.64}{$\textit{R}$}}}^2+{\mathcal{R}}\Pi_{{\scalebox{0.64}{$\textit{R}$}}}^\intercal(M_{{\scalebox{0.6}{$\textsc{P}$}}})=M^2_{{\scalebox{0.6}{$\textsc{P}$}}}$, where the ${\mathcal{R}}$ operator takes the real part of a function. Also, it yields the residue of the $B_\mu(x)$ propagator not to be exactly equal to unity. 

\section{The quantum transition amplitude}\label{quantum}

After deriving the full set of constraints and also choosing appropriate gauge conditions to turn the system into a second-class one, the next step is to obtain the transition amplitude for the quantum theory. This object is originally written in the phase space, and the measure should take into account the constraint structure of the model. Our goal is to use the phase space path integral for the vacuum-to-vacuum transition amplitude, within the Faddeev--Senjanovic method \cite{Faddeev:1969su,Senjanovic:1976br} to obtain such an object and, after some developments, to use a technique to turn the system into a covariant one. Subsequently, we will show that it can be written indeed in terms of the action as the majority of the usual gauge theories.
 The path integral formulation supplies a side of QFT that is non-perturbative, in principle. Hence, several techniques, from semiclassical approximations to numerical simulations, can be employed, which permit computing  non-perturbative effects. The path integral formulation also establishes deep connections between QFT and  critical phenomena. Even from the  computational viewpoint, there are circumstances  where  computations based upon the path integral setup are straightforward when compared to canonical quantization.
Consequently, as the canonical quantization and the path integral quantization  complement each other for Dirac fermions, the aim here consists of extending this intimate relationship to the case of mass dimension one fermions in the context of generalized quantum electrodynamics.

The mentioned phase space path integral for the vacuum-to-vacuum transition amplitude can be expressed as 
\begin{align}
Z=\!\!\!\bigintss\! \mathcal{D}_{A,B}\,\mathcal{D}\phi \,\mathcal{D} p_\phi\, \mathcal{D}\uppi\,\mathcal{D}\bar \uppi\, \mathcal{D}{\mathfrak{f}}\, \mathcal{D}{\gdualn{\mathfrak{f}}}\, \sqrt{\det{\left[\Omega_{jk}\right]}}\left[\prod_{\ell=1}^8\delta\left(\Sigma_\ell\right)\right]\exp\left(i\int d^4x\mathcal{L}_C\right),\label{eq11}
\end{align}
with the composed functional measure
\beq
\mathcal{D}_{A,B} = \prod_{\mu=0}^3\mathcal{D}A^\mu \prod_{\nu=0}^3\mathcal{D}B^\nu \prod_{\gamma=0}^3\mathcal{D}\Pi_\gamma^A\prod_{\xi=0}^3\mathcal{D}\Pi_\xi^B,
\eeq
and the canonical Lagrangian \begin{align}
\mathcal{L}_C=p_\phi\dot \phi+\Pi_\mu^A \dot{A}^\mu +\Pi_\mu^B \dot{B}^\mu+\gdualn{\uppi} \dot{{\mathfrak{f}}}+ \dot{{\gdualn{\mathfrak{f}}}}\uppi -\mathcal{H}_C.
\end{align}
Explicitly, Eq. (\ref{eq11}) reads
\beq
Z&=&\bigintss \mathcal{D}_{A,B}\mathcal{D}\phi\, \mathcal{D}p_\phi\, \mathcal{D}\uppi\,\mathcal{D}\bar \uppi\, \mathcal{D}{\mathfrak{f}} \mathcal{D}{\gdualn{\mathfrak{f}}} \det\big[(\nabla^2)^2\delta^{(3)}(\vec x-\vec y)\big]\left[\prod_{\ell=1}^8\delta\left(\Sigma_\ell\right)\right]\nonumber \\ 
&& \times\exp\left\{i\int d^4x\left[p_\phi \dot \phi+\Pi_\mu^A \dot A^\mu+\Pi_\mu^B \dot B^\mu+\gdualn{\uppi} \dot {\mathfrak{f}}+\dot {\gdualn{\mathfrak{f}}} \uppi- \left(\frac{1}{2}p^2_\phi\!-\!MB_0p_\phi\!+\! \gdualn{ \uppi} \uppi+ \nabla {\gdualn{\mathfrak{f}}}  \nabla {\mathfrak{f}}\right.\right.\right. \nonumber \\&&\left.\left.\left.\quad\quad\qquad\qquad+m^2{\gdualn{\mathfrak{f}}} {\mathfrak{f}} + \frac{1}{2}\Pi^A_i\Pi^A_i\!-\!\frac{1}{2}\Pi^B_i\Pi^B_i\!+\!\Pi^{i A}\partial_iA_0\!+\!\Pi^{i B}\partial_iB_0+ig\big(\Pi^{i A}\!-\!\Pi^{i B}\big) {\gdualn{\mathfrak{f}}} \sigma_{0i}{\mathfrak{f}}\right.\right.\right.\nonumber \\
&&\left.\left.\left.\quad\quad\qquad\qquad+\frac{1}{4}F_{ij}F_{ij}\!-\!\frac{1}{4}G_{ij}G_{ij}-\frac{ig}{2}{\gdualn{\mathfrak{f}}} \sigma_{ij} {\mathfrak{f}} \big(F_{ij}\!+\!G_{ij}\big)\!-\!\frac{M^2}{2}\!\left(\!B_i\!+\!\frac{\partial_i \phi}{M}\right)^2\right) \right]\right\},
\eeq
which can be simplified using the following identities
 \beq
 \delta\big(\partial_i\Pi^A_i\big)=\int \mathcal{D}\theta \exp{i\int d^4x\big(\theta \partial_i\Pi^A_i \big)},\eeq and \beq
 \delta\big(\partial_i\Pi^B_i-Mp_\phi\big)=\int \mathcal{D}\chi \exp\left[i\int d^4x \chi\big( \partial_i\Pi^B_i-Mp_\phi \big)\right].\eeq 
\indent The transition amplitude becomes
\beq
Z&=&\bigintss \mathcal{D}_{\vec A,\vec B}\,\mathcal{D}\phi\, \mathcal{D}p_\phi \, \mathcal{D} \theta\,  \mathcal{D} \chi\, \mathcal{D}{\mathfrak{f}}\, \mathcal{D}{\gdualn{\mathfrak{f}}}\, \det{[(\nabla^2)^2\delta^{(3)}(\vec x-\vec y)]}\left[\delta\left(\partial_iA_i\right)\right]\left[\delta\left(\partial_iB_i\right)\right]\nonumber \\
&&\times\exp \left\{\int d^4x\left(\left(\partial_\mu {\gdualn{\mathfrak{f}}} \partial^\mu {\mathfrak{f}}-m^2{\gdualn{\mathfrak{f}}} {\mathfrak{f}} \right)+p_\phi\dot \phi+\Pi^{i A} \dot A_i+\Pi^{i B} \dot B_i-\left[ \frac{1}{2}p^2_\phi+ \frac{1}{2}\Pi^A_i\Pi^A_i-\frac{1}{2}\Pi^B_i\Pi^B_i\right.\right.\right.\nonumber \\
&&\left.\left.\left.\quad\quad\qquad\qquad +\Pi_i^A\partial_i\theta +\Pi_i^B\partial_i\chi+Mp_\phi \chi
+\frac{1}{4}F_{ij}F_{ij}\!-\!\frac{1}{4}G_{ij}G_{ij}+ig\left(\Pi^{i A}-\Pi^{i  B}\right) {\gdualn{\mathfrak{f}}} \sigma_{0i}{\mathfrak{f}}\right.\right.\right.\nonumber \\
&&\quad\quad\qquad\qquad\left.\left.\left.-i\frac{g}{2}{\gdualn{\mathfrak{f}}} \sigma_{ij} {\mathfrak{f}} \left(F_{ij}+G_{ij}\right)-\frac{M^2}{2}\left(B_i+\frac{1}{M}{\partial_i\phi }\right)^2\right] \right)\right\}.\label{ta1}
\eeq
with the new definition
\beq
\mathcal{D}_{\vec A,\vec B} = \prod_{\mu=1}^3\mathcal{D}A^\mu \prod_{\nu=1}^3\mathcal{D}B^\nu \prod_{\gamma=1}^3\mathcal{D}\Pi_\gamma^A\prod_{\xi=1}^3\mathcal{D}\Pi_\xi^B.
\eeq
Considering the integration of the delta functions involving $A_0$ and $B_0$, these fields are substituted in the action by expressions associated with the interaction with DM and also the Stueckelberg field $\phi(x)$. However, these expressions are multiplied, up to integration by parts, by $\Sigma_3(x)$ and $\Sigma_4(x)$ defined in Eqs. (\ref{sig3}, \ref{sig4}),  which appear in the delta functions of the measure. Therefore, they do not contribute, and this specific part of the integration follows analogously to the free case. The next step is to use the functional identities and then associate $\theta$ with $A_0$ and $\chi$ with $B_0$.\\
\indent The transition amplitude (\ref{ta1}) is correct, but not explicitly Lorentz covariant. To turn the system into a covariant one, the following identity with a covariant gauge condition should be considered \cite{m2y}  \bea  \mathfrak{I}=\int D\alpha(x) \det [\Box \delta^{\left(4\right)}\left(x-y)\right] \delta \Big(  \partial^\mu A_\mu^\alpha- f(x)  \Big),                  \eea
with $A_\mu^\alpha$  being the gauge-transformed field
\bea A_\mu(x)\mapsto A_\mu(x)+\partial_\mu \alpha(x),  \eea
and $f(x)$ being a scalar field. A similar expression regarding the $B_\mu(x)$ field and its associated gauge freedom related to the parameter $\Lambda(x)$ is introduced. This identity can be inserted into the path integral, the inverse gauge transformation can be performed, as the system is gauge invariant, and then we can integrate with respect to $\alpha$ to eliminate the presence of $\det \nabla^2 \delta^{\left(3\right)}\left(\vec x-\vec y\right) $  and $\delta\big(\partial_iA_i\big)$ as well, from the integration measure. After this procedure, and  integrating over the different scalar functions $f$, $\int df {\cal{M}}(f)=1$ with \beq{\cal{M}}(f)=N\exp\left(-\frac{i}2\int d^4x {\xi f^2(x)}\right),\eeq 
with a normalization factor $N$.\\
 \indent Applying an analogous procedure for $B_\mu(x)$, it yields
\beq Z&=&N\int \prod_{\mu=0}^3\mathcal{D}A^\mu \prod_{\nu=0}^3\mathcal{D}B^\nu \det{[(\Box) \delta^{(4)}(x-y)]}\det{[(\Box) \delta^{(4)}(x-y)]}\,\mathcal{D}\phi\, \mathcal{D}{\gdualn{\mathfrak{f}}}\, \mathcal{D} {\mathfrak{f}}\nonumber \\&&\times\exp\left[i\int d^4x\left( -\frac{1}{4}F_{\mu \nu}F^{\mu \nu}-\frac{1}{2} \left[M^2\left(B_\mu+\frac{1}{M}\partial_\mu \phi\right)^2+B^\mu\Box \theta_{\mu \nu}B^\nu  \right]\right. \right. \nonumber \\&&\left.\left.\qquad\quad\qquad\qquad+\partial_\mu {\gdualn{\mathfrak{f}}} \partial^\mu {\mathfrak{f}}-m^2{\gdualn{\mathfrak{f}}} {\mathfrak{f}}+\frac{ig}{2}{\gdualn{\mathfrak{f}}} \sigma_{\mu \nu} {\mathfrak{f}} \left(F^{\mu \nu}+G^{\mu \nu}\right)\right.\right.\nonumber \\ &&\left.\left.\qquad\quad\qquad\qquad+\frac{\Omega}{2}\big(\partial_\mu B^\mu\big)^2+\frac{\xi}{2}\big(\partial_\mu A^\mu \big)^2 +J_\mu A^\mu+K_\mu B^\mu +\bar T {\mathfrak{f}}+{\gdualn{\mathfrak{f}}} T+J\phi \right)\right],     \label{eqq}    \eeq
where $\det{[\Box \delta^{(4)}(x-y)]}^2$ can be absorbed by a normalization factor or can be written in terms of a set of two decoupled ghost actions associated to the two independent {U}(1) local symmetries of the model. This last discussion, of course, is just valid for the present case with $T=0$. In the last step, sources for each quantum field were added.

\subsection{Schwinger--Dyson and the complete quantum equations }\label{quantum1}

 {The most elegant way of studying the set of field equations  is a functional formulation, consisting
of an infinite chain of coupled differential equations, relating to different Green functions. The infinite tower of equations is referred to as the Schwinger-Dyson  equations that will be addressed in what follows. This non-perturbative approach will give us a phenomenological sense of the complete DM-photon quantum interaction structure in the electrodynamics investigated, due to the skeleton diagrams, and thus pave the way for a perturbative analysis by the study of the radiative corrections \cite{Ro,Robert1,Robert2,Albino,Nash,Landau}.} It is worth mentioning that this non-perturbative approach has important applications in several fields such as QCD$_4$ and also planar reduced QED in condensed matter context, for example \cite{Albino}.

\subsubsection{The Schwinger--Dyson equations for the \textcolor{black}{fermionic} propagator}

The expression to the full fermion propagator can be derived starting from Schwinger variational equation
\begin{equation}
\left[ \left. \frac{\delta S}{\delta\gdualn{\mathfrak{f}}\left( x\right) }\right\vert
_{ \frac{\delta}{\delta iT } ,\frac{\delta}{\delta i\bar{T}},\frac{\delta}{\delta iJ } ,\frac{\delta}{\delta  iJ_{\mu } },\frac{\delta}{\delta  iK_{\mu } }} +\bar{T}\left( x\right) %
\right] Z\left[ T ,\bar{T},J, J_{\mu }, K_{\nu}\right] =0 . \label{var schw dkp}
\end{equation}%
 Then, expressing Eq.  \eqref{var schw dkp} in terms of the connected Green function generator $\mathcal{W}$ and then differentiating the resulting expression with respect to the source $T \left( y\right)$, yields 
\beq
i\delta^{\left(4\right) }\!\left(x\!-\!y\right) \!&=&\!-\!\left[(\Box+m^{2})\!+\!\frac{ig}{2}\sigma_{\mu \nu}\left\langle F^{\mu \nu}\!+\!G^{\mu\nu}\right\rangle\!+\!\frac{ig}{2}\sigma_{\mu \nu}F^{\mu \nu}\!\left(\frac{\delta}{\delta  iJ_{\mu }}\right)\right.\nonumber\\&&\qquad\qquad\qquad\qquad\qquad\qquad\qquad\qquad\left.+\!\frac{ig}{2}\sigma_{\mu \nu}G^{\mu\nu}\!\left(\frac{\delta}{\delta  iK_{\mu } }\right)\right] \mathcal{S}\left(x,y;A,B\right), \label{eq4}
\eeq
where the vacuum expectation values of the field strengths in Eq. (\ref{eq4}) have been taken into account.
The functional   \beq -i\mathcal{S}(x,y;A,B)\equiv \frac{\delta^2 \mathcal{W}}{\delta T(y)\delta \bar{T}(x)}\Bigg\vert_{A,B}\eeq represents the complete DM propagator in the case of nonzero vacuum averages. Working out the field variations, we can identify
\begin{equation}
\Sigma \left(x,z\right) =\lim_{y \to x}ig^{2}\sigma^{\mu \gamma }\partial_\gamma^{y}\left( \int d^{4}ud^{4}v \,\mathcal{S}\left( x,v;A,B\right) \Gamma _{\alpha }\left( v,z;u\right)\mathfrak{D}%
_{\mu }^{\alpha } \left( u,y ;A,B\right)\right).\label{sigma}
\end{equation}%
The distribution
 \beq \mathfrak{D}%
_{\mu }^{\alpha } \left( x,y ;A,B\right)\equiv \frac{\delta^2 \mathcal{W}}{\delta J_\alpha(y)\delta J^\mu(x)}\Bigg\vert_{A,B}\eeq 
is the generalized Podolsky propagator in the presence of non-zero vacuum averages, being a difference between the complete Maxwell and the complete Proca 2-point functions, whose inverses are going to be defined in the next subsections.  \\
The definition of the complete vertex function is given by
\bea g\Gamma ^{\xi}_{pq}(x,w,y) \equiv  \frac{\delta^3\Gamma}{\delta {\gdualn{\mathfrak{f}}}_p(w)\delta {\mathfrak{f}}_q(y)\delta  A_{\xi}(x)},\eea
which, as it is going to be proven, is equal to the vertex function  calculated through  $B_\mu(x)$ field variations.

Taking the limit of vanishing field vacuum averages yields
\begin{equation}\label{1234}
i\delta ^{\left( 4\right) }\left( x-y\right) =-\left[(\Box+m^2)\right] \mathcal{S}\left( x-y\right) +\int d^{4}z\Sigma \left( x-z\right) \mathcal{S}\left(z-y\right),
\end{equation}
considering that at the limit, a specific coordinate dependence is assumed due to translation symmetry. The $\mathcal{S}(x-y)$ operator denotes the complete propagator, whereas the $S(x-y)$ distribution indicates the bare propagator. In the momentum representation Eq. (\ref{1234}) becomes  \beq\mathcal{S}^{-1}\left( p\right) =S^{-1}(p)+\Sigma \left( p\right).\label{1235}\eeq
Eq. (\ref{1235}) can be viewed in Fig. \ref{full}.   The double bosonic internal line denotes the sum of the massive and the massless propagators.

Eq. \eqref{1234} can be formally  written as
\begin{equation}
\mathcal{S}\left( p\right) =\frac{i}{p^{2}-\mathfrak{M} \left( p\right)},  \label{propag}
\end{equation}%
where the mass operator $\mathfrak{M}$ reads 
\begin{equation}
\mathfrak{M}(p)=m^{2}+\Sigma (p).  \label{mass operator}
\end{equation}

\subsubsection{The Schwinger--Dyson equations for the photon propagator}

The complete expression of the gauge-field propagator can be determined through the functional generator leading to the Schwinger variational equation for the gauge field,
\begin{equation}
\left[ \left. \frac{\delta \mathit{S}}{\delta A_{\gamma }\left( x\right) }%
\right\vert _{ \frac{\delta}{\delta iT } ,\frac{\delta}{\delta i\bar{T} } ,\frac{\delta}{\delta  iJ_{\mu } },\frac{\delta}{\delta  iK_{\mu } },\frac{\delta}{\delta  iJ }} +J^{\gamma}\left( x\right) \right] Z\left[ T ,\bar{T},J, J_{\mu }, K_{\nu}\right] =0.
\end{equation}
In terms of the functional $Z\left[ T ,\bar{T},J, J_{\mu }, K_{\nu}\right] =\exp \left\{ i\mathcal{W}%
\left[ T ,\bar{T},J, J_{\mu }, K_{\nu}\right] \right\} $ for the connected Green functions, we obtain the equation 
\beq
-J^{\gamma }\left(x\right) &=&ig\,Tr\left[%
\sigma ^{\gamma \nu}\partial_\nu \left( \frac{\delta^2 \mathcal{W}}{\delta T\left(x\right)\delta \bar{T}\left( x\right) }%
\right)\right] +ig   Tr\left[\sigma^{\gamma \nu}\partial_\nu^x \Big(\frac{\delta \mathcal{W}}{\delta T\left( x\right) }\frac{%
\delta \mathcal{W}}{\delta \bar{T}\left( x\right) }\Big) \right]\nonumber\\
&&+\left[\theta^{\gamma \mu }+\xi\ l^{\gamma \mu }\right]\square \frac{\delta \mathcal{W}}{\delta J^{\mu }\left(x\right)}.
\label{func green conectadas}
\eeq
Eq. \eqref{func green conectadas} can be interpreted as the complete photon field equation subjected to an external source $J^{\gamma }$. In this equation $%
\theta^{\gamma \mu }$ and $l^{\gamma \mu }$ are differential projectors given by 
\begin{equation}
\theta^{\gamma \mu }+l^{\gamma \mu }=\eta^{\gamma \mu }, \qquad\qquad l^{\gamma \mu }=\frac{\partial ^{\gamma }\partial ^{\mu }}{\square }.
\end{equation}%

To obtain the complete gauge-field propagator, it proves convenient to introduce also the generating
functional for the one-particle irreducible (1PI)\footnote{When it is not possible to compose a disconnected Feynman diagram  by merely cutting one line, the resulting graph is called one-particle irreducible (1PI). Instead, when it is possible to split a Feynman diagram into two disjoint parts by cutting a unique line, a one-particle reducible graph is obtained, which does not represent new divergences.} Green functions, which is related to $\mathcal{W}$ by a functional Legendre transformation 
\beq\Gamma=\mathcal{W}-\int d^4x\Big(J_\mu A^\mu+K_\mu B^\mu +\gdualn{T} {\mathfrak{f}}+{\gdualn{\mathfrak{f}}}T  +J\phi \Big).\label{1PI}\eeq 
Hence, rewriting \eqref{func green conectadas} in terms of the 1PI functional and varying it with relation to $A_\nu(y)$ yields
\begin{align}
\frac{\delta ^{2}\Gamma  }{\delta
A_{\nu }\left( y\right) \delta A_{\gamma }\left( x\right) }=&\left[ \theta^{\gamma \nu }+\xi\ l^{\gamma \nu }\right]\square \delta ^{\left( 4\right) }\left( x-y\right) \nonumber \\
&+ig^{2}Tr\left[\sigma^{\mu \gamma}\partial_\mu^x \int d^{4}ud^{4}w\Big( \mathcal{S}\left( x-u\right)  \Gamma ^{\nu }\left( u,w;y\right)\mathcal{%
S}\left( w-x\right) \Big) \right]
\label{tchegando}
\end{align}
at the limit of vanishing field averages.\\
\indent Analogously, for the massive gauge boson, the following relation holds,
\begin{align}
\frac{\delta ^{2}\Gamma  }{\delta
B_{\nu }\left( y\right) \delta B_{\gamma }\left( x\right) }=&-\left[ \left(\theta^{\gamma \nu }+\Omega\ l^{\gamma \nu }\right)\Box+M^2\eta^{\gamma \nu}\right] \delta ^{\left( 4\right) }\left( x-y\right) \nonumber \\
&+ig^{2}Tr\left[\sigma^{\mu \gamma}\partial_\mu^x \int d^{4}ud^{4}w\Big( \mathcal{S}\left( x-u\right)  \Gamma ^{\nu }\left( u,w;y\right)\mathcal{%
S}\left( w-x\right) \Big) \right].
\label{massive}\end{align}
 Both 2-point functions are associated with the inverse of the massless and the massive gauge field complete propagators, respectively. 
The second term of Eq. \eqref{tchegando} can be identified with the photon self-energy tensor, $\Pi^{\gamma \nu }$,
\begin{equation}
\Pi^{\gamma \nu }\left( x,y\right) =ig^{2}Tr\Big(\sigma^{\gamma \mu}\partial_\mu^x \int d^{4}ud^{4}w\left[
\mathcal{S}\left( x,u\right)  \Gamma ^{\nu }\left( u,w;y\right)\mathcal{S}\left(
w,x\right) \right]\Big). \label{autoenergia foton1}
\end{equation}%
 The transverse nature of this complete object is going to be inferred in the subsection regarding the Ward--Takahashi-like identities.\\
 \indent It is also important to derive the following result
 \begin{align}
\frac{\delta ^{2}\Gamma  }{\delta
B_{\nu }\left( y\right) \delta A_{\gamma }\left( x\right) }=
&+ig^{2}Tr\left[\sigma^{\mu \gamma}\partial_\mu^x \int d^{4}ud^{4}w\Big( \mathcal{S}\left( x-u\right)  \Gamma ^{\nu }\left( u,w;y\right)\mathcal{%
S}\left( w-x\right) \Big) \right],
\label{massive1}\end{align}
demonstrating that radiative corrections generate mixing between massless and massive fields. This point is going to be further addressed in the last section about renormalization. In Fig. \ref{full}, the simple bosonic line may denote the massive, the massless, and also the inverse of the radiatively generated mixed $AB$ propagator. $P^{-1}$ is a general notation for the bare contribution for each of these cases.

\subsubsection{Schwinger--Dyson equation for the vertex part}

\begin{figure}[h!]
\centering
\includegraphics[width=16cm]{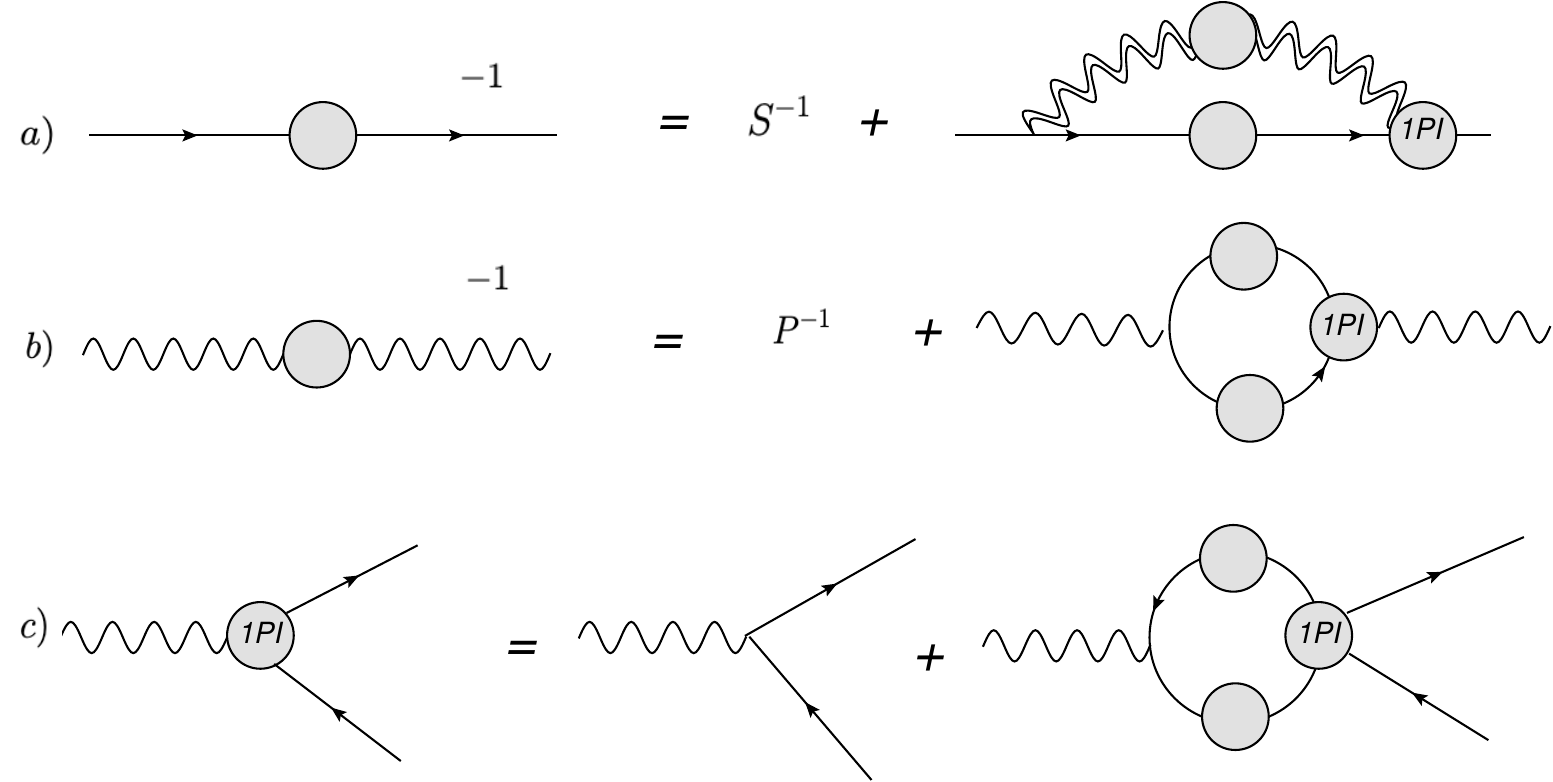} 
\caption{Full propagators and vertex of the quantum electrodynamics investigated. Double lines of photons represent the sum of the contribution of the massive and the massless bosons.}
\label{full}
\end{figure}

\indent To derive the expression for the vertex function, we write  Eq. \eqref{func green conectadas}  in terms of the quantum action and perform field variations with respect to the fermionic fields. We are going to demonstrate that this trilinear function is the same when calculated via massless or  massive field variations, since it depends just on the fermionic 1PI structures. In the derivation of the fermionic and bosonic self-energies,   this property was implicitly considered. Then, discarding fermion number symmetry violating terms, it yields\footnote{In this expression, the Latin letters denote spinor indexes.}
\beq
g\Gamma ^{\xi}_{pq}(x,w,y) &=&ig\sigma^{\xi \nu}_{pq}\partial_\nu^x\Big(\delta^4(x-w)\delta^4(x-y)\Big)\nonumber\\
\qquad\qquad&&\qquad\qquad 
+ig^2\sigma_{bd}^{\xi \nu}\partial_\nu^x\Big[ \int d^4z\ d^4t\ \mathcal{S}_{da}(x-t) \Phi_{pqca}(w,y,z,t) \mathcal{S}_{cb}(z-x)        \Big],\label{vert}
\eeq
with the definition
\begin{eqnarray}g\Phi_{pqca}(w,y,z,t)\equiv  \frac{\delta^4\Gamma}{\delta {\gdualn{\mathfrak{f}}}_p(w)\delta {\mathfrak{f}}_q(y)\delta \mathfrak{f}_c(z) \delta   {\gdualn{\mathfrak{f}}}_a(t)}.\label{the end} \end{eqnarray}
From Eqs. \eqref{vert}  and \eqref{the end} the 3-point vertex function can be seen to depend on the 4-point fermion-fermion one. Diagrammatically, the irreducible vertex part can be visualized in Fig. \ref{full}.

\subsubsection{Schwinger--Dyson equation for Compton 4-point function}

\begin{figure}[h!]
\centering
\includegraphics[width=16cm]{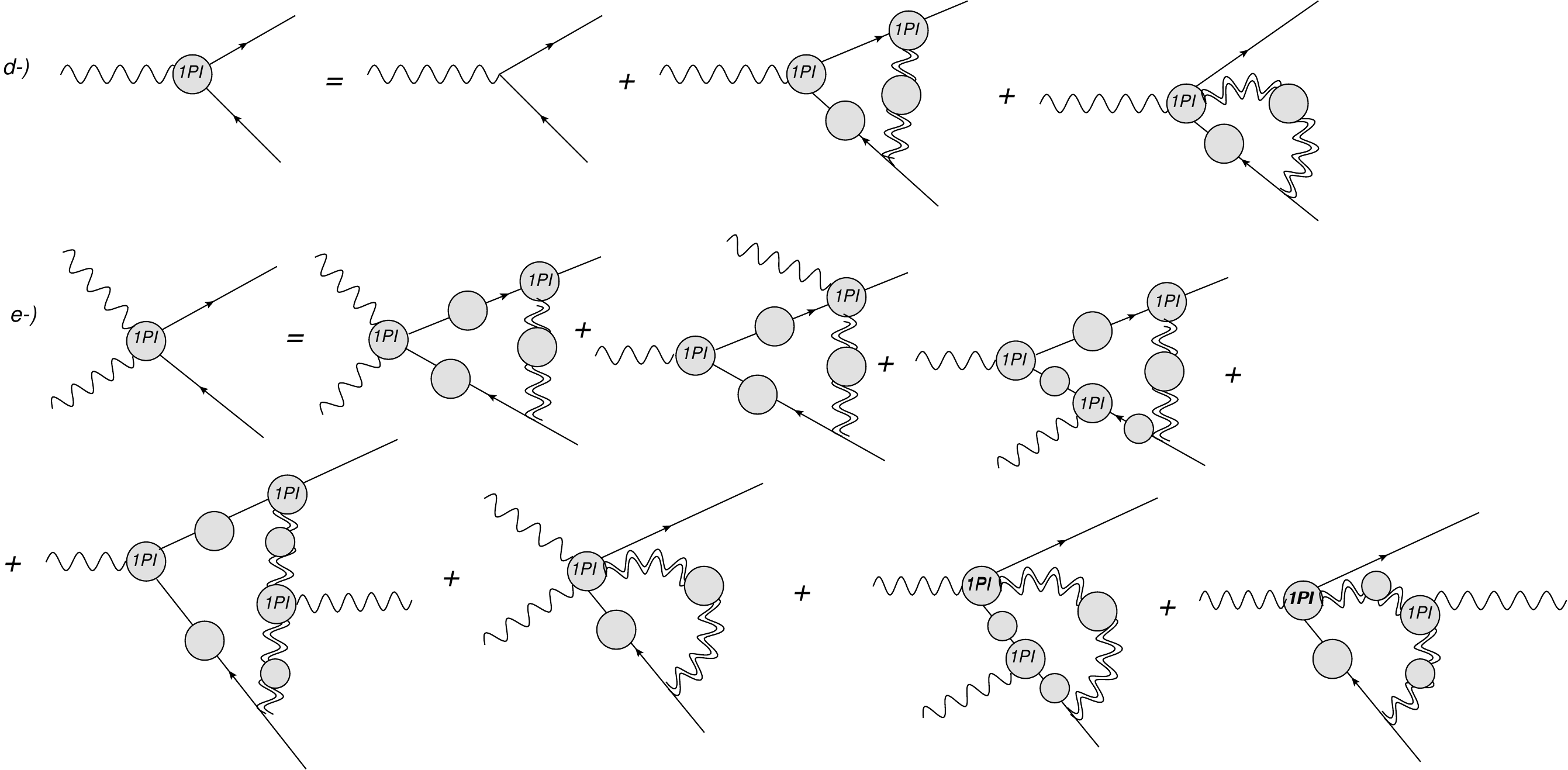} 
\caption{Full vertex as a function of 4-point Compton function and full 4-point Compton function for the electrodynamics investigated.}
\label{full_4_point_compton}
\end{figure}

To gain familiarity with the Schwinger--Dyson chain associated with the investigated electrodynamics, let us now apply  the functional variation with respect to the  $A_{\sigma}$ gauge field in Eq. (\ref{eq4}). It yields the following result,
\beq
\!\!\!\!&&\!\!\!\!\frac{i\delta ^{3}\Gamma }{\delta A_{\sigma }(z)\delta\gdualn{\mathfrak{f}}_{a}(x)\delta{\mathfrak{f}}_{b} (y)}=g\sigma^{\sigma \nu}_{ab}\partial_\nu^x\delta ^{4}(x-z)\delta ^{4}(y-x)\nonumber \\ 
\!\!\!\!&&\!\!\!\!+ig\sigma^{\sigma \nu}_{ac}\partial_\nu^x\bigintsss\!\! d^{4}w\frac{\delta ^{2}\Gamma }{\delta\gdualn{\mathfrak{f}}_{c}(x)\delta{\mathfrak{f}}_{d} (w)}\frac{\delta }{\delta J^{\mu }(x)}\left[\bigintsss d^{4}ud^{4}t\frac{%
\delta ^{2}\mathcal{W}}{\delta\gdualn{T}_{d}(w)\delta{T}_{e} (u)}\frac{\delta ^{3}\Gamma }{%
\delta A_{\sigma }(z)\delta\gdualn{\mathfrak{f}}_{e}(u)\delta{\mathfrak{f}_{m}} (t)}\frac{\delta ^{2}\mathcal{W}}{\delta\gdualn{T}_{m}(t)\delta{T}_{b} (y)}\right]\nonumber \\ 
\!\!\!\!&&\!\!\!\!+ig\sigma^{\sigma \nu}_{ac}\partial_\nu^x\bigintsss\!\! d^{4}w\frac{\delta ^{2}\Gamma }{\delta\gdualn{\mathfrak{f}}_{c}(x)\delta{\mathfrak{f}}_{d} (w)}\frac{\delta }{\delta K^{\mu }(x)}\left[\bigintsss d^{4}ud^{4}t\frac{%
\delta ^{2}\mathcal{W}}{\delta\gdualn{T}_{d}(w)\delta{T}_{e} (u)}\frac{\delta ^{3}\Gamma }{%
\delta A_{\sigma }(z)\delta\gdualn{\mathfrak{f}}_{e}(u)\delta{\mathfrak{f}}_{m} (t)}\frac{\delta ^{2}\mathcal{W}}{\delta\gdualn{T}_{m}(t)\delta{T}_{b}(y)}\right].\nonumber\\\!\!\!\! 
\eeq
After simplifications, an alternative vertex structure
can be attained, 
\beq
&&\frac{i\delta ^{3}\Gamma }{\delta A_{\sigma }(z)\delta\gdualn{\mathfrak{f}}_{a}(x)\delta{\mathfrak{f}}_{b} (y)}=g\sigma^{\sigma \nu}_{ab}\partial_\nu^x\delta ^{4}(x-z)\delta ^{4}(y-x)\nonumber\\&&
+ig\sigma^{\sigma \nu}_{ac}\partial_\nu^x
\left\{\bigintsss d^{4}s\frac{\delta ^{2}\mathcal{W}}{\delta J^{\rho }(s)\delta J^{\mu }(x)}%
\left[\bigintsss d^{4}t\left(\frac{\delta ^{4}\Gamma }{\delta A_{\rho }(s)\delta A_{\sigma
}(z)\delta\gdualn{\mathfrak{f}}_{c}(x)\delta{\mathfrak{f}}_{d} (t)}\right)\frac{\delta ^{2}\mathcal{W}}{\delta\gdualn{T}_{d}(t)\delta{T_{b}(x)}}\right.\right.\nonumber\\&&
\left.\left.-\bigintsss d^{4}ud^{4}t\bigintsss d^{4}r\frac{\delta ^{3}\Gamma }{\delta A_{_{\rho
}}(s)\delta\gdualn{\mathfrak{f}_{c}}(x)\delta{\mathfrak{f}}_{d}(r)}\frac{\delta ^{2}\mathcal{W}}{\delta\gdualn{T}_{d}(r)\delta{T}_{e}(u)}\frac{\delta ^{3}\Gamma }{\delta A_{\sigma
}(z)\delta\gdualn{\mathfrak{f}}_{e}(u)\delta{\mathfrak{f}}_{m}(t)}\frac{\delta ^{2}\mathcal{W}}{\delta\gdualn{T}_{m}(t)\delta{T}_{b}(y)}\right]\right\}\nonumber\\&&
+ig\sigma^{\sigma \nu}_{ac}\partial_\nu^x\left\{\bigintsss d^{4}s\frac{\delta ^{2}\mathcal{W}}{\delta K^{\rho }(s)\delta K^{\mu }(x)}%
\left[\bigintsss d^{4}t\left(\frac{\delta ^{4}\Gamma }{\delta B_{\rho }(s)\delta A_{\sigma
}(z)\delta\gdualn{\mathfrak{f}}_{c}(x)\delta{\mathfrak{f}}_{d} (t)}\right)\frac{\delta ^{2}\mathcal{W}}{\delta\gdualn{T}_{d}(t)\delta{T_{b}(y)}}\right.\right.\nonumber\\&&
\left.\left.-\bigintsss d^{4}ud^{4}t\bigintsss d^{4}r
\frac{\delta ^{3}\Gamma}{\delta B_{\rho}(s)\delta\gdualn{\mathfrak{f}}_{c}(x)\delta{\mathfrak{f}}_{d} (r)}\frac{\delta^{2}\mathcal{W}}{\delta\gdualn{T}_{d}(r)\delta{T}_{e}(u)}\frac{\delta ^{3}\Gamma }{\delta A_{\sigma
}(z)\delta\gdualn{\mathfrak{f}}_{e}(u)\delta{\mathfrak{f}}_{m} (t)}\frac{\delta ^{2}\mathcal{W}}{\delta\gdualn{T}_{m}(t)\delta{T}_{b}(y)}\right]\right\}.\nonumber\\
\label{comptoneq}
\eeq
Now, the definitions for $\Gamma$, ${\cal S}$ and ${\cal D}$, and the 4-point function associated with the Compton effect  
\begin{equation}
\frac{\delta ^{4}\Gamma }{\delta A_{\rho }(s)\delta A_{\sigma }(z)\delta\gdualn{\mathfrak{f}}_{a}(t)\delta{\mathfrak{f}}_{b}(y)}\dot{=}g^{2}\Phi ^{\rho \sigma }_{ab}(t,y;z;s)
\end{equation}%
yield
\beq
i\Gamma ^{\sigma }_{ab}(x,y;z)&=&\sigma^{\sigma \nu}_{ab}\partial_\nu^x\delta ^{4}(x-z)\delta ^{4}(y-x)\delta ^{4}(x-z)\delta
^{4}(y-x)\nonumber\\&&
+ig^{2}\sigma^{\mu\nu}_{ac}\partial_\nu^x\bigintsss d^{4}sd^{4}ud^{4}td^{4}r{\cal D}_{\mu \rho
}(y,s)\Gamma ^{\rho }_{cd}(r,x;s){\cal S}_{de}(u,r)\Gamma ^{\sigma }_{em}(t,u;z){\cal S}_{mb}(x,t)\nonumber\\&&
-ig^{2}\sigma^{\mu\nu}_{ac}\partial_\nu^x\bigintsss d^{4}sd^{4}t{\cal D}_{\mu \rho }(x,s)\Phi ^{\rho \sigma
}_{cd}(t,x;z;s){\cal S}_{db}(y,t).
\label{v3}
\eeq
The previous result can be represented, for the vertex function related to the Compton 4-point function, in Fig. \ref{full_4_point_compton}. Interestingly, the possibility of two equivalent expressions for the vertex yields a non-perturbative relation between two whole classes of graphs composed of an infinite number of diagrams. This is analogous to the QED$_4$ case. The only difference is the presence of double internal bosonic lines,  denoting the superposition of massive and massless propagators, reflecting the structure of our model. This second equivalent expression is useful for developing first-order corrections for the vertex.

Finally, varying Eq. (\ref{comptoneq}) with respect to the  $A_{\theta}(w)$ gauge  field yields
\begin{eqnarray}
&&\frac{i\delta ^{4}\Gamma }{\delta A_{\theta}(w)\delta A_{\sigma }(z)\delta\gdualn{\mathfrak{f}}_{a}(x)\delta{\mathfrak{f}}_{b} (y)}\cr\cr
&=&ig\sigma^{\sigma \nu}_{ac}\partial_\nu^x\left\{\bigintsss  d^{4}s\frac{\delta ^{3}\mathcal{W}}{\delta A_{\theta}(w)\delta J^{\rho }(s)\delta J^{\mu }(x)}\bigintsss  d^{4}t\left(\frac{\delta ^{4}\Gamma }{\delta A_{\rho }(s)\delta A_{\sigma
}(z)\delta\gdualn{\mathfrak{f}_{c}}(x)\delta{\mathfrak{f}}_{d} (t)}\right)\frac{\delta ^{2}\mathcal{W}}{\delta\gdualn{T}_{d}(t)\delta{T_{b}(y)}}\right.\nonumber\\&&\left.+
\bigintsss d^{4}s\frac{\delta ^{2}\mathcal{W}}{\delta J^{\rho }(s)\delta J^{\mu }(x)}%
\bigintsss  d^{4}t\left(\frac{\delta ^{5}\Gamma }{\delta A_{\theta}(w)\delta A_{\rho }(s)\delta A_{\sigma
}(z)\delta\gdualn{\mathfrak{f}_{c}}(x)\delta{\mathfrak{f}}_{d} (t)}\right)\frac{\delta ^{2}\mathcal{W}}{\delta\gdualn{T}_{d}(t)\delta{T_{b}(y)}}\right.\nonumber\\
&&\left.+\bigintsss\! d^{4}s\frac{\delta ^{2}\mathcal{W}}{\delta J^{\rho }(s)\delta J^{\mu }(x)}\!
\bigintsss\! d^{4}t\left(\frac{\delta ^{4}\Gamma }{\delta A_{\rho }(s)\delta A_{\sigma
}(z)\delta\gdualn{\mathfrak{f}_{c}}(x)\delta{\mathfrak{f}}_{d} (t)}\right)\!\frac{\delta ^{3}\mathcal{W}}{\delta A_{\theta}(w)\delta\gdualn{T}_{d}(t)\delta{T_{b}(y)}}\!+\!\cdots\right\}
\end{eqnarray}
wherein the ellipsis represents all the other terms due to the photon field variations that could be obtained directly, and we see that the 4-point Compton function depends on a five-point function.  We can synthesize the result, considering all the insertion of external photons in the previous vertex structure of this subsection, as seen in Fig. \ref{full_4_point_compton}. This object furnishes a useful tool for the discussion of our concept of darkness in Sec. \ref{compton}. Namely, the evaluation of the 1-loop corrections for Compton/pair annihilation processes.

\subsection{Ward--Takahashi-like identities and the quantum gauge symmetry}\label{quantum2}
When one considers extending the derivation of Noether’s theorem, implementing a variation of the field comprising a symmetry of the Lagrangian leads to a classically conserved current. A similar variation on the path integral yields a resourceful relationship among correlation functions, known as the Ward--Takahashi identities. They are an implication of the gauge invariance and consist of a   non-perturbative technique, playing a  prominent role in the renormalization of the DM-photon interaction.
An interesting observation regarding the problematic issues of a non-gauge fixed local symmetric theory can be readily noted in the expression (\ref{eqq}). In the absence of the gauge fixing term for the massive vector field $B_\mu(x)$, the redefinition  $B_\mu(x) \mapsto B_\mu(x)-\frac{1}{M}\partial_\mu \phi(x)$ leads to an explicit divergent functional generator due to the integration in $\mathcal{D}\phi(x)$. This is just another piece of evidence that the addition of the gauge fixing term is essential to have a well-defined path integral.
 Varying the functional generator with respect to an infinitesimal gauge symmetry transformation gives \begin{align} \!\!\!\!\!\!\!\!\delta Z\big[ \bar{T},T,J,K_\mu,J_\mu\big ]\!=\!\!\bigintssss\!\! \prod_{\mu=0}^3\mathcal{D}A^\mu\! \prod_{\nu=0}^3\mathcal{D}B^\nu \mathcal{D}\phi(x)\mathcal{D}{\gdualn{\mathfrak{f}}} \mathcal{D} {\mathfrak{f}}  \! \int\! \!d^4x\Big(\xi \Box(\partial_\mu A^\mu)\!+\!  \partial_\mu J^\mu \Big)\alpha(x)\! \exp\!\left(i\!\!\int d^4x \mathcal{L}_{{\scalebox{0.6}{$\textsc{pod}$}}}\!\right).  \end{align}
 The notation  ${\mathcal{L}}_{{\scalebox{0.6}{$\textsc{pod}$}}}$ is  understood as the previously defined Lagrangian (\ref{pod2}) plus source terms.

The variation of the measure has been not considered, since the Jacobian due to this infinitesimal gauge transformation is unitary. Expressing the gauge invariance of the functional generator in terms of the connected Green function generator, defined by $Z=e^{i\mathcal{W}}$, yields the following expression
\begin{align} &\int d^4x\left(\partial^\mu J_{\mu}(x)+\xi \Box \partial^\alpha \frac{\delta \mathcal{W}}{\delta J_{ \alpha}(x)} \right)Z=0. \end{align}
\indent Considering the expression for the effective action \cite{Peskin:1995ev},  a relevant relation can be derived, which depends just on the fields and the $\Gamma$ variations with respect  to them, 
\begin{align}&\partial^\mu \frac{\delta \Gamma}{\delta A_\mu(x)}+\xi \Box (\partial_\mu A^\mu(x))=0.\label{f}\end{align}
 Varying this expression with respect to the gauge field, taking the longitudinal projection, and considering the limit of all fields going to zero at the end of the calculations, we can obtain an expression whose momentum space version reads
\bea \xi k^4+k_\gamma k^\mu \frac{\delta^2 \Gamma}{\delta A_\gamma(k)\delta A^{\mu}(-k)}=0,\eea 
where the translation symmetry of the theory, at the vanishing classical field limit, was considered. The  renormalized fields are defined as
\beq\label{epp}\Phi_A(x)=\sqrt{Z}_{\Phi} \Phi_A^{{\scalebox{0.64}{$\textit{R}$}}}(x)+\cdots
\eeq 
\noindent and the theory's  parameters as $\mathfrak{Y}=Z_\mathfrak{Y} \mathfrak{Y}^{{\scalebox{0.64}{$\textit{R}$}}}$, with $\Phi_A(x)$ and $\mathfrak{Y}$ being collective designations for all the fields and parameters, respectively, also defining  $Z_{\mathfrak{Y}}=1+\delta_{\mathfrak{Y}}$.
The ellipsis in Eq. (\ref{epp}) denotes the renormalization mixing to be applied here, in which just the diagonal terms contribute to the  longitudinally projected Ward--Takahashi-like  identities. Therefore, $Z_\xi Z_A=1$, at all orders.

A completely analogous procedure for the local symmetry associated with $B_\mu$ yields
\bea \Omega k^4+k_\gamma k^\mu \frac{\delta^2 \Gamma}{\delta B_\gamma(k)\delta B^{\mu}(-k)}=iMk_\mu \frac{\delta^2 \Gamma}{\delta B_\mu(-k) \delta \phi(k) }.\eea 
\indent The 2-point function relating the massive gauge field and the auxiliary scalar one is going to be developed in the next paragraphs. Then, it is used in advance here and proved later. It yields the following non-perturbative constraint on the longitudinal part of the gauge field 2-point function
\bea \Omega k^4+k_\gamma k^\mu \frac{\delta^2 \Gamma}{\delta B_\gamma(k)\delta B^{\mu}(-k)}=M^2k^2.\label{1236}\eea 
Renormalizing Eq. (\ref{1236}) leads to $Z_\Omega Z_{B}=1$ and also $Z_MZ_B=1$.
These results imply that the boson self-energy tensor $\Pi_{\mu \nu}(k)$ satisfies  \beq
k^\mu \Pi_{\mu \nu}(k)=0.\eeq Since no longitudinal terms arise from radiative corrections, it implies that\footnote{The mass term is written in terms of the Minkowski metric ${\eta}_{\mu \nu}$ which has a transverse and a longitudinal contribution ${\eta}_{\mu \nu}=\theta_{\mu \nu}+\omega_{\mu \nu}$. We are considering the renormalization $M^2=Z_M M^2_{{\scalebox{0.64}{$\textit{R}$}}}$.} $Z_MZ_{B}=1$ for the full quantum corrections, reinforcing this previous conclusion. Here, $Z_M$ is associated with Podolsky mass renormalization. Indeed, the massive term has projections on both the transverse and the longitudinal projectors. 

The Stueckelberg pure gauge particle is regulated by the following quantum equation of motion,
\bea \frac{\delta \Gamma}{\delta \phi(k)}=-M\left(ik^\nu B_\nu(x)+\frac{1}{M}k^2 \phi(k)\right).            \eea
Field variations imply that  
\bea \frac{\delta^2 \Gamma}{\delta \phi(k)\delta \phi(-k)}=-k^2 , \qquad \qquad  \frac{\delta^2 \Gamma}{\delta \phi(k)\delta B_\nu(-k)}=-iM k_\nu .       \eea
 Interestingly, it is decoupled, up to longitudinal interactions. Considering that the interaction is transverse and this longitudinally coupled particle is defined a priori to be out of the physical spectrum, then there is no interaction to renormalize it,  $Z_\phi=1$, and the relation between $Z_M$ and $Z_B$ is kept the same, demonstrating the internal consistency of the theory. These are the only non-vanishing quantum action variations involving the Stueckelberg field.
 
\indent Now, implementing field variations on Eq. (\ref{f}), to derive Ward--Takahashi-like  identities, yields 
\begin{align}k_\alpha \frac{\delta^3\Gamma}{\delta {\gdualn{\mathfrak{f}}}(p)\delta {\mathfrak{f}}(\tilde p)\delta  A_{\alpha}(k)}=0.       \end{align}
 Considering for $B_\mu(x)$ the same approach concerning its associated gauge symmetry also leads to
\begin{align}k_\alpha \frac{\delta^3\Gamma}{\delta {\gdualn{\mathfrak{f}}}(p)\delta {\mathfrak{f}}(\tilde p)\delta  B^{\alpha}(k)}=0.      \end{align}
If we continue to implement field variations on Eq. (\ref{f}), to derive Ward--Takahashi-like  identities, we will have, for example, the transverse condition for the Compton 4-point function
\begin{eqnarray}
&&k_\alpha \frac{\delta^4\Gamma}{\delta {\gdualn{\mathfrak{f}}}(p)\delta {\mathfrak{f}}(\tilde p)\delta  A_{\beta}(k')\delta  A_{\alpha}(k)}=0=k_\alpha \frac{\delta^4\Gamma}{\delta {\gdualn{\mathfrak{f}}}(p)\delta {\mathfrak{f}}(\tilde p)\delta  B_{\beta}(k')\delta  B^{\alpha}(k)},
\end{eqnarray}
on which is natural to apply the previous identity in the study of the Compton scattering.

\begin{figure}[h!]
\centering
\includegraphics[width=7cm, height=9cm]{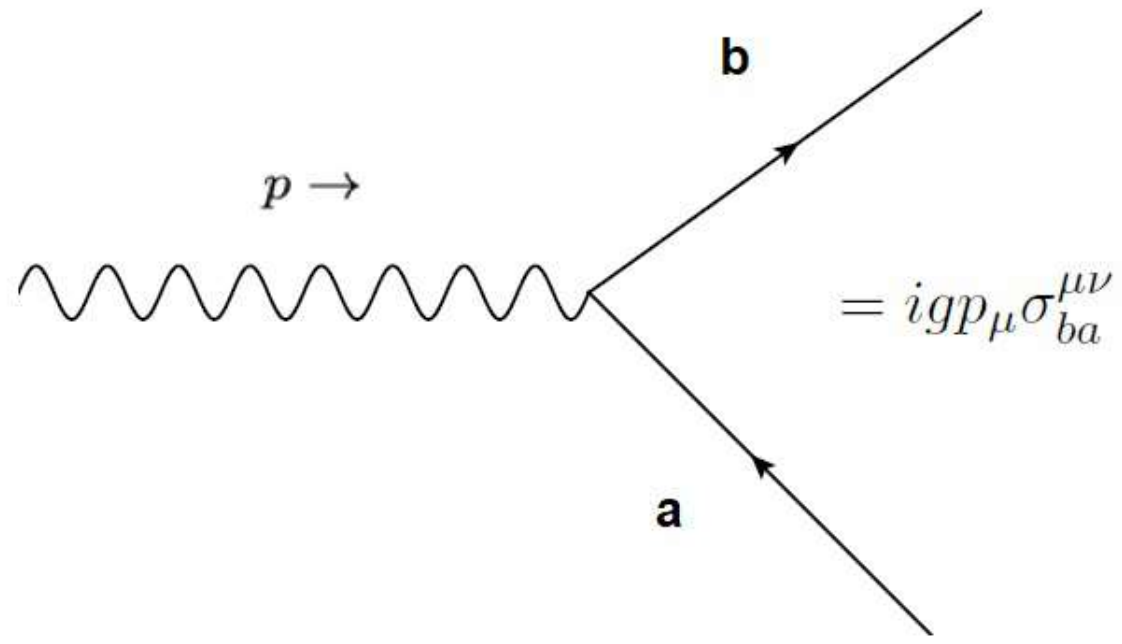}
\caption{Vertex of the generalized electrodynamics investigated.}
\label{vertex}
\end{figure}

\begin{figure}[h!]
\centering
\includegraphics[width=14cm]{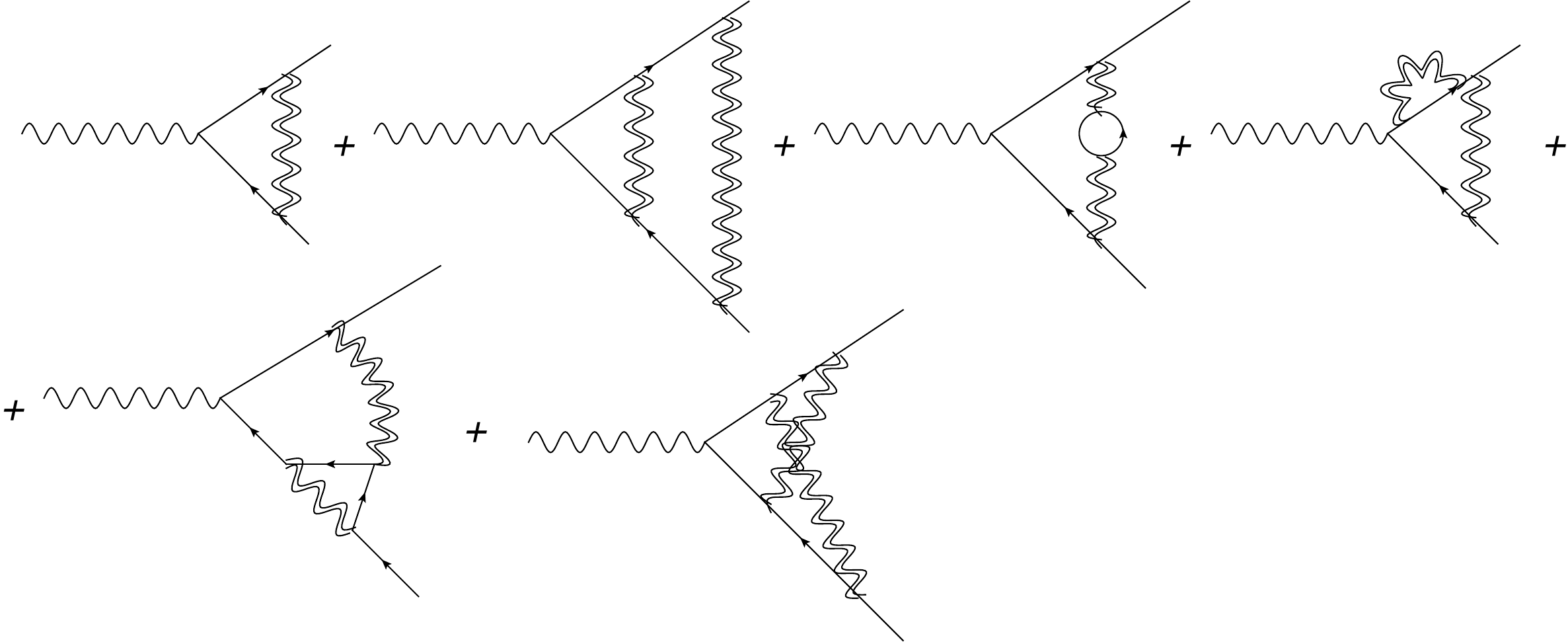}
\caption{Vertex radiative corrections.}
\label{vertexradi}
\end{figure}

The explicit lowest order diagram for the vertex is a key ingredient for our discussion and can be inferred by the Schwinger--Dyson equation \eqref{comptoneq}. Its structure is displayed in Fig. \ref{vertex}.
 Taking it into account, this equation can be iteratively solved to yield the full set of corrections for the vertex up to two loops. See Fig. \ref{vertexradi}.

In order to analyze this content, let us first remember that a basis for the Clifford--Dirac algebra of the $4\times 4$ gamma matrices can be obtained by the set 
\bea \mathfrak{G}_a=\left\{\mathbb{I}_{4\times 4}, \gamma_\mu, \gamma_\mu \gamma_5,\gamma_5, \sigma_{\mu \nu}\right\},  \eea
with a norm defined as $\langle A,B\rangle =\frac{1}{4}Tr(A^\dagger B)$ leading to $\langle  \mathfrak{G}_a, \mathfrak{G}_b\rangle=\delta_{ab}.$ 

This knowledge can be used to obtain the general form for the vertex function until $2$-loop order. Namely, considering that the vertex structure of Fig. \ref{vertex} implies that until the 2-loop setup of Fig. \ref{vertexradi}, the 3-point function is traceless in the spinor indexes, and owing to the explicit form of the 1-loop radiative corrections highlighted in the final section,  the following expression can be inferred, 
\bea  \frac{\delta^3\Gamma^{\textsc{2-loop}}}{\delta {\gdualn{\mathfrak{f}}}(p)\delta {\mathfrak{f}}(\tilde p)\delta  A_{\alpha}(k)}\equiv ig\Gamma_\alpha(p,\tilde p,k=p-\tilde p)=\ ig\mathfrak{F}(p,\tilde p,k=p-\tilde p)\sigma_{\nu \alpha }k^\nu.         \eea
valid at least at this order. The term  $\mathfrak{F}(p,\tilde p,k=p-\tilde p)$ denotes an arbitrary function, dependent on its explicit arguments.  Besides the  explicit structure of the 1-loop corrections, it was also considered that the complete vertex is composed of an even number of gamma matrices and that   the trace of its product with $\gamma_5$ should be zero. One can argue that such a trace, in four dimensions, should be written in terms  of the antisymmetric Levi--Civita tensor. Then, to build a transverse Lorentz vector through this tensor and the three external momenta, one faces an impossible task,  since they are constrained. These properties, the knowledge of the spinor basis,  and the fact that the lowest order propagators are proportional to the identity are enough to achieve this result. For higher corrections, the transverse constraint is enough just to guarantee that this vertex can be written in terms of a linear combination of $12$ operators \cite{Ro,Robert1,Robert2}.

 \indent Differently from the well-known standard QED$_4$ case, in which the fermion self-energy depends on the gauge and just the structure with attached external fermionic lines is gauge-independent, our bare vertex is identically transverse and the associated complete self-energy is gauge-invariant. This is a consequence of the fact that our model couples directly with the physical fields $\vec E(x)$ and $\vec B(x)$. The fact that this feature is kept under radiative is due to the Ward identity.  Expressing the previously obtained relation for the complete fermion propagator as
 \bea \mathcal{S}^{-1}(x-s)=i\big(\Box+m^2\big)\delta^4(x-s)+ig^2\int d^4z\ d^4u\ \sigma^{\mu \beta}\mathcal{S}(x-u)\Gamma_\nu(u,s,z)\partial^x_\beta \big[\mathfrak{D}_\mu^{\ \nu}(z-x)\big] \eea
 one conclude that since the bare vertex and, as we have proved, the complete dressed one is identically transverse, no gauge parameter dependence due to longitudinal terms of the complete bosonic propagator occurs. Their association with the gauge parameters and the fact that they do not receive quantum corrections is ensured by the Ward--Takahashi-like identities. In other words, even considering the full radiative corrections, \textcolor{black}{fermionic mass dimension one} quantum field keeps being coupled just to the physical fields and not directly to the gauge field  \cite{Ahluwalia:2022ttu}.

\section{On the non-polarized Compton-like and pair annihilation processes }
\label{compton}

\indent  Our definition of darkness is associated with zero amplitude for Compton and pair annihilation unpolarized processes at $1$-loop and low scattering angles. 
Interestingly, using Mandelstam variables, the scattering amplitude for the tree-level Compton-like process can be written as 
\bea
    {\cal{M}}_{\upalpha,\upalpha^\prime,\upbeta,\upbeta^\prime}&=&\frac{g^2}{2m\big(s-m^2\big)}i\gdualn \varepsilon_\upbeta(p')\upnu^{ \nu}(k')\upnu^{ \sigma}(k)\upepsilon_\nu^{*\upalpha}(k') \upepsilon^{\,\upalpha'}_\sigma(k)\varepsilon_{\upbeta'}(p)\nonumber \\
    &&+\frac{g^2}{2m\big(u-m^2\big)}i\gdualn \varepsilon_\upbeta(p')\upnu^\sigma(k)\upnu^{\nu}(k')\upepsilon_\nu^{*\upalpha}(k') \upepsilon^{\ \upalpha'}_\sigma(k)\varepsilon_{\upbeta'}(p).
\eea
The notation used is  standard. Therefore, $\upepsilon^{\,\upalpha}_\sigma$ denotes the external photon with polarization $\upalpha$ and $\upnu_\mu(k)=\gamma_\mu \slashed{k}-k_\mu$ here is assumed to be the bare vertex. 
Replacing the $s$ channel by $t$, $\gdualn \varepsilon_\upbeta(p')$ by  $\gdualn \chi_\upbeta(p')$ and with all external photons being conjugated, we obtain the amplitude for the DM pair annihilation. The $2m$ in the denominator is related to the Feynman rules \cite{Duarte:2020svn,Alves:2017joy}. Since  the massive intermediate boson is a Merlin mode, and, therefore, it does not contribute to the external states and just the massless one enters these calculations \cite{Donoghue:2021eto},  we  consider $k^2={k'}^2=0$.  The normal  resonance propagates  forward in time, with positive energy, whereas this massive
pole propagates backward,  distinguishing the Merlin mode  from the ghost, which has just a minus sign in the numerator appearing in  the propagator. 
Faddeev--Popov ghosts present a negative sign in the numerator, however, they carry the usual imaginary unit, in the denominator. In the case of the Merlin mode, a change of sign occurs in the denominator as well,  yielding the associated propagator to represent the time-reversed version
of the standard resonance propagator. It does not violate unitarity, but just microcausality in a  timescale  $\Delta t\sim M_{{\scalebox{0.6}{$\textsc{P}$}}}/ \upgamma$, with $\upgamma$ being  the imaginary part of the total photon self-energy tensor, the sum of each kind of interaction contributions, evaluated at the propagator pole mass.  If it has the opposite sign, this contribution grows indefinitely, leading to ill behavior. However, in our case, it decays exponentially. Additionally, considering the experimental constraints, $g$ is expected to be small, and the dominant contribution to photon self-energy tensor comes from the interaction with Dirac fermions of generalized QED$_4$.\\
\indent Regarding the smallness of the coupling constant $g$, 
 the LUX and the XENON1T experiments   impose an upper bound on DM/nucleon cross-section of order $10^{-45}\,{\rm cm}^2$, for a DM mass  $m \sim 10\ {\rm GeV}$ \cite{LUX:2016ggv,XENON:2017vdw}. More recently, LUX-ZEPLIN experiment  found this same order of magnitude for the cross-section at this mass \cite{LZ:2022ufs}. However, for the specific case of spin-dependent DM/proton interaction, it presents a less stringent limit $\sigma \lesssim 10^{-39} {\rm cm}^2$ at $m\approx 10 \ {\rm GeV}$. We consider this order for $m$ since it is the most kinematically favorable for xenon devices. There are very stringent limits for $m\gtrsim30 \ {\rm GeV}$ \cite{Aalbers:2022dzr}, and also due to the recently obtained low energy excess for $\sim 5\ {\rm GeV}$ DM mass in precise  simulations  regarding energy losses for several phonon-mediated detection experiments \cite{Sassi:2022njl}. 
 
 Although we are mainly focusing on the   DM-generalized photon interaction, to obtain a good estimate of $\Delta t$ the non-idealized situation with all the concomitantly occurring interactions must be considered.  The mentioned microcausality violation occurs, considering the 1-loop dominant contribution from generalized QED$_4$, in a time interval $\Delta t\sim \frac{1}{e^2M_{{\scalebox{0.5}{$\textsc{P}$}}}}$ with the  bound on the mass  $M_{{\scalebox{0.6}{$\textsc{P}$}}} \gtrsim 370$ GeV \cite{Bufalo:2014jra} and the electric charge being  $e\approx 0.302$, in natural units,  leading to $\Delta t \sim 10^{-25}s$. More specifically, we consider the massive excitations from the Podolsky model as virtual particles due to their small lifetimes, unitarity compatibility, and microcausality violation. Then, it contributes just to internal lines. 
Interestingly, this upper bound on the massive excitation lifetime is of the order of the $Z$ boson, one of the smallest of the standard model. Differently from the latter, this Merlin mode propagates backward, and then, no physical decay can be associated with it. Otherwise, it would contribute to generating pair production signatures that are, in principle, measurable. Then, the best way to deal with such causality violation is indeed to consider the massive excitation as a virtual particle, complying with the conceptual basis of quantum field theory. 
\\
\indent The non-polarized squared amplitude calculated through the $\ddagger$ twisted  conjugation 
 \bea \sum_{\upalpha,\upalpha^\prime,\upbeta,\upbeta^\prime}\!\!\!{\cal{M}}_{\upalpha,\upalpha^\prime,\upbeta,\upbeta^\prime}{\cal{M}}^\ddagger_{\upalpha,\upalpha^\prime,\upbeta,\upbeta^\prime}\label{tww} \eea 
is proportional to a linear combination of the following terms
\bea    Tr[\upnu^\nu(k)\upnu^\sigma(k')\upnu_\sigma(k')\upnu_\nu(k)], \qquad\qquad\qquad  \quad    Tr[\upnu^\sigma(k')\upnu^\nu(k)\upnu_\sigma(k')\upnu_\nu(k)],       \eea
for the highlighted Compton amplitude and also for the mentioned pair annihilation process. 
The use of the twisted  conjugation in Eq. (\ref{tww}) is necessary to yield a quantum field theory compatible with the generalized optical theorem. The latter is associated with the $S$-matrix unitarity in the context of the $\ddagger$ twisted conjugation  prescription.

\begin{figure}[h!]
\centering
\includegraphics[width=15cm]{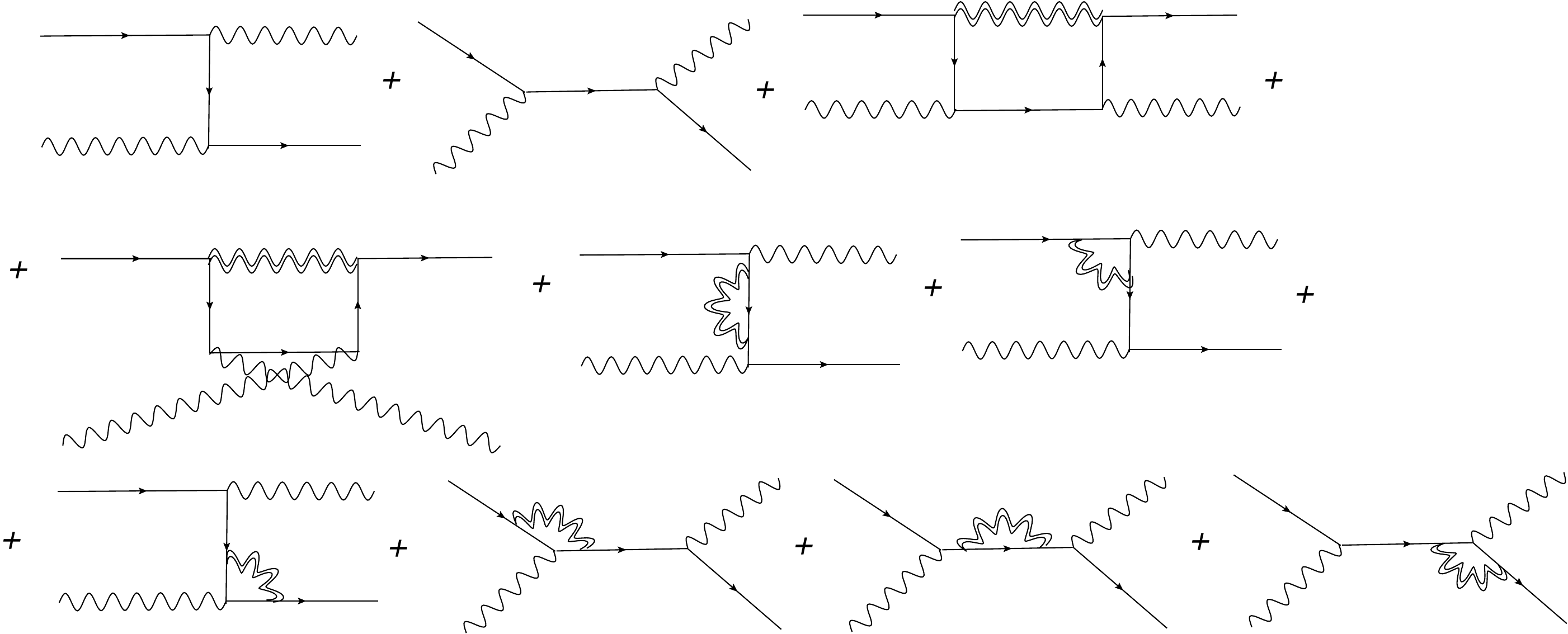}
\caption{Compton radiative corrections.}
\label{comptonradi}
\end{figure}

Both traces are proportional to $\left(k^\mu k_\mu\right)( k'_\mu {k'}^{\mu})$ meaning, according to our discussion, that the probability for these non-polarized processes involving external light vanishes. We call it darkness property in strong form.  If one considers, beyond the tree level, the next-order corrections associated with the box diagram, arising from iterations on the Schwinger--Dyson equation \eqref{v3}, new classes of contributions must be considered. There are also corrections due to the replacement of the  propagator, the vertexes, and the parameters with their complete versions over the tree-level structure. The sum of these terms is displayed in Fig. \ref{comptonradi}.

One important example of these corrections is the 4-point correlation box diagram associated with the quantum action variations concerning two \textcolor{black}{fermions} and two photons \eqref{v3}. The fundamental ingredients are the propagators and the vertex. The mentioned darkness property is violated if one considers an arbitrary number of loop corrections. Apart from the four external field box, this feature is still valid for the remaining 1-loop corrections.  However, the total squared amplitude has some terms that explicitly violate it.  \\
\indent The box diagram, which violates the strong darkness property, can be obtained by the lowest-order iteration of the Schwinger--Dyson \eqref{v3}. The box diagram is proportional to the term 
\bea  M^2g^4\int \frac{d^4l}{(2\pi)^4}\frac{\upnu^\alpha(l)\left(\upnu^\mu(k)\upnu^\nu(k')   \right)\upnu_\alpha(l)}{\big[(p-l)^2-m^2\big]\big[(q-l)^2-m^2\big]\big[(p'-l)^2-m^2\big]\big[l^2-M^2\big]\big[l^2\big]},\label{num19} 
\eea
with $q=p+k$. Here, $p$ and $k$ denote the incoming \textcolor{black}{fermionic mass dimension one field} and photonic momentum, respectively. The $p'$ and $k'$ represent their outgoing versions. The crossed term with interchanged bosonic lines is obtained by replacing $k$ by $k'$ and also $\nu$ by $\mu$ indexes. The numerator in Eq. (\ref{num19}) can be managed to yield
\bea     -l^\beta l_\beta\left(\upnu^\mu(k)\upnu^\nu(k')  \right)+\slashed{l}\gamma_\alpha\left(\upnu^\mu(k)\upnu^\nu(k')   \right)\gamma^\alpha \slashed{l}.\eea 
Therefore, using the spinor basis and the identity $\gamma_\alpha \sigma_{\mu \nu}\gamma^\alpha=0$, the second term can be expressed as 
\bea      l^\sigma l_\sigma \Big\{  -4i\big(\epsilon^{\mu \nu \alpha \beta }k_\alpha' k_\beta\big)\  \gamma_5+4\big[k_\mu k_\nu'     -\eta_{\mu \nu} (k\cdot k') \big] \mathbb{I}_{4\times 4}           \Big \}.                      \eea
It clearly shows that the massless pole cancels out, leading to better infrared behavior. This is an important feature since the weak darkness principle is associated with the low-energy limit. Regarding the non-polarized squared amplitude, the   crossed terms between the tree level and the remaining 1-loop terms with the box diagrams vanish as $k_\mu k^\mu$. It can be readily verified,  if one uses the gamma matrices identity $\gamma_\mu \gamma_\alpha \gamma_\beta \gamma^\mu=2\eta_{\alpha \beta}$ as well as the fact that the vertexes commute if they are sandwiched by two contracted gamma matrices.  Although products between box diagrams violate this property, it must be highlighted that, regarding the squared amplitude, strong darkness is still valid until order $g^6$. As already  mentioned, this is an extremely small value, ensuring darkness at very high precision. The $g^8$ order can be further explored, as will be discussed now. \\
\indent There is also a weaker condition that is valid for all graphs at 1-loop. The proper amplitudes vanish for small enough scattering angles. In the center-of-mass reference frame, the incoming and outgoing photons are related by a rotation. For small angles, such that $ \sin\theta \sim \theta$,  if the amplitude can be factorized with one of the factors being the product of the vertexes associated with the external photons, it is possible to verify that it vanishes. Then, considering this setting, after attaching external fields to build the amplitudes and considering the transverse nature of the polarization vectors, the following factor in the numerator of all the tree level and $1$-loop Compton/pair annihilation processes arises
\bea     \slashed{\upepsilon}^{\,\upalpha}(k)  \slashed{k}  \slashed{\upepsilon}^{\,\upalpha'}(k')  \slashed{k'}=  -\slashed{\upepsilon}^{\,\upalpha}(k)  \slashed{k}\slashed{k'}    \slashed{\upepsilon}^{\,\upalpha'}(k')  = - k_\mu k^\mu \slashed{\upepsilon}^{\,\upalpha}(k)    \slashed{\upepsilon}^{\,\upalpha'}(k')= 0,                      \eea 
since, for low scattering angles,   an infinitesimal rotation matrix  
$ \mathfrak{H}_{\mu \nu}=\eta_{\mu \nu}+\Omega_{[\mu \nu]}$, with the second term being an antisymmetric tensor relates $k_\mu'$ and $k_\mu$.  It implies that this factor is proportional to $k^\mu k_\mu$ and must vanish. This approximation is valid for small angles  in the center-of-mass reference frame. Although $\theta$ is not the angle in the lab frame, for low momentum transfer, both tend to zero. Therefore, considering that at cosmological scales it is legitimate to consider low energy background photons and low momentum transfer, both the tree level and the 1-loop properties ensure darkness. It is associated with a precision of $g^{8}$ for its weaker form. This is our phenomenological definition.\\
\indent {Finally, it is worth mentioning the extremely low probability of experimentally measuring box diagrams contributions to the physical processes amplitudes at fourth order in the coupling, just exalting, for example, the very weak interaction between two photons \cite{David}} for the case of standard QED$_4$. Just to have an idea, the experiment built to detect them obtained $70$ events of a total of more than a hundred billion tentatives. Regarding our specific case, the boxes appear in one-loop computations for  Compton and pair annihilation-like processes. Taking into account the fact that nucleon recoil experimental bounds imply in the $g$ coupling is much smaller than electric charge, one can be easily convinced that our prescription is indeed enough to ensure darkness. 

We proved that our approach can describe  nucleon recoil processes without violating this DM-defining property.
The Schwinger--Dyson  equation for the \textcolor{black}{dark matter particle} propagator indicates that this one-loop condition is fulfilled. However, to trust this result, one must be sure that when renormalizing the theory no counterterm that does not appear in the bare Lagrangian must be included. Otherwise, the Schwinger--Dyson equation would need to be reconsidered, to take it into account. Hence, not only a structure whose divergence does not grow with the order of the correction must be assured, but also some other aspects. For example, although scalar QED is renormalizable, as can be computed and encoded into the superficial degree of divergence, it demands the inclusion of the interaction $\lambda \phi^4$ and also a new renormalization condition to have a non-singular 4-point function. Therefore, these aspects must be further investigated, to check whether a new interaction must be considered,  to ensure an ultraviolet completion.\\


\section{Accounting for  renormalizability and superrenormalizability}
\label{secre}
The renormalizability of the theory  describing DM-photon interaction in a Podolskyan generalized QED$_4$ 
can be then analyzed. 
The Feynman diagrams present an integration over $d^4q$ for each loop in the graph and a factor of the type $1/((q - p)^2 - m^2)$ for each fermionic propagator on an internal line, with $p$ denoting a  combination of external momenta. Therefore,  each loop integration takes into account  four powers of momenta at the numerator and each internal fermionic line two powers at the denominator. Regarding our model, it is important to obtain its superficial degree of divergence.
In general, any Feynman diagram contains the integral $\int
G(p,\ldots)\, d^4p$, where $G(p,\ldots)$ encodes the contributions with powers of $p$ coming from vertexes, propagators, and any other possibility for the associated Feynman diagram. The superficial degree of divergence, $D$, is defined by the ultraviolet (UV)  behavior of the
$\int
G(p,\ldots)\, d^4p\stackrel{{\rm UV}}{\sim} \int p^{D-4}\,d^4p$, 
given by \beq
D=4L+V-N_\gamma-2P_\mathfrak{f}-4P_\gamma,\eeq where $L$ is the number of loops and $N_\gamma$ denotes the number of external photon lines, whereas $P_\mathfrak{f}$ and $P_\gamma$ denote, respectively, the number of  fermion and photonic propagators.  The
number of loops can be expressed as
$L = N_i - V + 1$, where $V$ is the number of vertexes of the graph. The term $\big(V-N_\gamma \big) $ expresses the fact that since the vertex momentum is associated with the bosonic line when it is internal, it leads to an increase of the divergence degree, while when it is external, there is no such increase.  The coefficient $4$ for the bosonic line is due to the use of the Podolsky model since all the associated internal lines are the difference between the massive and the massless propagator. On the other hand, the factor $2$ for the \textcolor{black}{dark matter particle} is due to the fermionic second-order Lagrangian.
Since the topology of the graphs is the same as the usual QED${}_4$, the number of  vertexes of a Feynman diagram is given by  
 \beq
 V=2P_\gamma +N_\gamma=P_\mathfrak{f}+\frac{1}{2}N_\mathfrak{f},\eeq
 whereas the number of loops reads \beq
 L=P_\mathfrak{f}+P_\gamma-V+1.\eeq   From these constraints, we can conclude that  $D=4-N_\gamma-\frac{3}{2}N_\mathfrak{f}-P_\mathfrak{f}$. From the topological constraints, we have $P_\mathfrak{f}=V-\frac{1}{2}N_\mathfrak{f}$. Therefore, 
 \bea D=4-N_\gamma -N_\mathfrak{f}-V. \eea 
If Feynman diagrams contain hidden 2- or 4-point correlation functions with 1-loop (or more), they will diverge, despite the superficial degree of divergence.  According to Weinberg's theorem, any Feynman diagram converges if its superficial degree of divergence, together with the degree of divergence of all its subgraphs, is negative.
 \begin{figure}[h!]
\centering
\includegraphics[width=16.5cm]{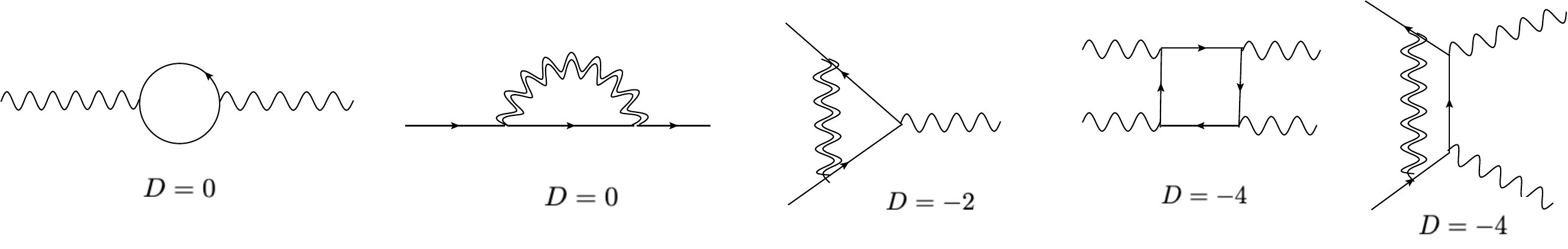}
\caption{Superficial degree of  divergences for processes with the usual photon and massive photon (combined $A+B$). Since the model is superrenormalizable, the graphs are labeled by their most divergent contribution, the 1-loop case. }
\label{div}
\end{figure}
 
 In our case, superrenormalizability can be read off the fact that the more vertexes are added, the more convergent the diagram is. In a superrenormalizable theory, only a finite number of Feynman diagrams superficially diverge. In this specific case, just the 1-loop boson and fermion self-energies are divergent. Some relevant graphs are displayed in Fig. \ref{div}. 
 It is worth mentioning that although the standard QED${}_4$ with Dirac fields is superrenormalizable only in less than four dimensions \cite{Peskin:1995ev},

 \subsection{The renormalization procedure}
 
 \indent Before proceeding with the renormalization, it is easier to first decouple the Stueckelberg field by 
 \bea B_\mu(x) \mapsto B_\mu(x)-\frac{1}{M}\partial_\mu \phi(x), \eea
 followed by the redefinition
 \bea \Box \phi(x) \mapsto \Box \phi(x)+M\partial_\mu B^\mu(x). \eea
 It then implies a decoupled $\phi(x)$ action whose integration leads to a differential operator-dependent normalization for the path integral. As mentioned, when deriving the path integral, normalization factors of this kind were obtained. They could be associated with the ghosts of the massless and the massive fields. The operator-dependent normalization that would arise from the implicit ghosts associated with U(1) symmetry of the massive field is canceled by the one due to Stueckelberg field integration.\\
 \indent Although it is not important for this specific Abelian case in $T=0$ analysis, the thermal properties are sensitive to this specific kind of normalization. Therefore, this is the well-defined procedure to correctly eliminate the Stueckelberg field, as well as the artificial gauge fixing terms associated with the massive boson.  Since the model is composed of two bosonic fields interacting with the fermionic mass dimension one field, the radiative correction generates a mixture of these terms. Hence,  to renormalize the system, the following mixing must be considered\footnote{Here $\theta_{\mu \nu}=\eta_{\mu \nu}+\partial_\mu \partial_\nu \frac{1}{\Box}$ is a formal notation to be understood in the Green function sense. It is worth mentioning that it leads to a local renormalized theory.} \cite{Martellini:1997mu}
 \begin{subequations}\beq A_\mu(x) &=& \sqrt{Z_A}A^{{\scalebox{0.64}{$\textit{R}$}}}_\mu(x)+\tilde \delta\ \theta_\mu^{\ \nu} B^{{\scalebox{0.64}{$\textit{R}$}}}_\nu(x),\\
 B_\mu(x)&=& \sqrt{Z_B}\mathfrak{F}_{{\scalebox{0.64}{$\textit{R}$}}}B^{{\scalebox{0.64}{$\textit{R}$}}}_\mu(x),\eeq
 \end{subequations} 
 leading to the following renormalized quantum Lagrangian
 \beq  \!\!\!\!\!\! {\cal{L}}^{{\scalebox{0.6}{$\textsc{quant}$}}}\!&=&\!-\frac{1}{4}F_{\mu \nu}^{{\scalebox{0.64}{$\textit{R}$}}}F^{\mu \nu}_{{\scalebox{0.64}{$\textit{R}$}}}-\frac{1}{2} \left[M^2_{{\scalebox{0.64}{$\textit{R}$}}}\left(\mathfrak{F}_{{\scalebox{0.64}{$\textit{R}$}}}B_\mu^{{\scalebox{0.64}{$\textit{R}$}}}\right)^2+\mathfrak{F}_{{\scalebox{0.64}{$\textit{R}$}}}^2B^{\mu}_{{\scalebox{0.64}{$\textit{R}$}}}\Box \theta_{\mu \nu}B^{\nu}_{{\scalebox{0.64}{$\textit{R}$}}} \right] +\big(\partial_\mu {\gdualn{\mathfrak{f}}}^{{\scalebox{0.64}{$\textit{R}$}}} \partial^\mu {\mathfrak{f}}^{{\scalebox{0.64}{$\textit{R}$}}}-m^2{\gdualn{\mathfrak{f}}}^{{\scalebox{0.64}{$\textit{R}$}}} {\mathfrak{f}}^{{\scalebox{0.64}{$\textit{R}$}}} \big)\nonumber\\ &&+\frac{\xi_{{\scalebox{0.64}{$\textit{R}$}}}}{2}\left(\partial^\mu A_\mu^{{\scalebox{0.64}{$\textit{R}$}}}\right)^2+\frac{i}{2}{\gdualn{\mathfrak{f}}}^{{\scalebox{0.64}{$\textit{R}$}}} \sigma_{\mu \nu} {\mathfrak{f}}^{{\scalebox{0.64}{$\textit{R}$}}} \big(g_A^{{\scalebox{0.64}{$\textit{R}$}}}F^{\mu \nu}_{{\scalebox{0.64}{$\textit{R}$}}}+g_B^{{\scalebox{0.64}{$\textit{R}$}}}G^{\mu \nu}_{{\scalebox{0.64}{$\textit{R}$}}}\big)+\tilde \delta  A_\mu^{{\scalebox{0.64}{$\textit{R}$}}}\Box \theta^{\mu \nu}B_\nu^{{\scalebox{0.64}{$\textit{R}$}}} -\frac{\delta_A}{4}F_{\mu \nu}^RF^{\mu \nu}_{{\scalebox{0.64}{$\textit{R}$}}}\nonumber \\
  && -\frac{1}{2} \left[\left(\mathfrak{F}_{{\scalebox{0.64}{$\textit{R}$}}}^2\delta_B-{\tilde{\delta}}^2\right) B^{\mu}_{{\scalebox{0.64}{$\textit{R}$}}}\Box \theta_{\mu \nu}B^{\nu}_{{\scalebox{0.64}{$\textit{R}$}}} \right] +\left(\partial_\mu {\gdualn{\mathfrak{f}}}^{{\scalebox{0.64}{$\textit{R}$}}} \partial^\mu {\mathfrak{f}}^{{\scalebox{0.64}{$\textit{R}$}}}\delta_2-m^2_{{\scalebox{0.64}{$\textit{R}$}}}\left(\delta_m+\delta_2\right){\gdualn{\mathfrak{f}}}^{{\scalebox{0.64}{$\textit{R}$}}} {\mathfrak{f}}^{{\scalebox{0.64}{$\textit{R}$}}} \right)\nonumber\\&&+\frac{i}{2}{\gdualn{\mathfrak{f}}}^{{\scalebox{0.64}{$\textit{R}$}}} \sigma_{\mu \nu} {\mathfrak{f}}^{{\scalebox{0.64}{$\textit{R}$}}} g_A^{{\scalebox{0.64}{$\textit{R}$}}}\left(\delta_{g_A}+\frac{\delta_A}{2}+\delta_2\right)F^{\mu \nu}_{{\scalebox{0.64}{$\textit{R}$}}}+g_B^{{\scalebox{0.64}{$\textit{R}$}}}\frac{i}{2}{\gdualn{\mathfrak{f}}}^{{\scalebox{0.64}{$\textit{R}$}}} \sigma_{\mu \nu} {\mathfrak{f}}^{{\scalebox{0.64}{$\textit{R}$}}}\left(\delta_{g_B}+\tilde{\delta}+\frac{\delta_B}{2}+\delta_2\right)G^{\mu \nu}_{{\scalebox{0.64}{$\textit{R}$}}}\nonumber \\ 
   &&+\frac{1}{2}\left(g_A^{{\scalebox{0.64}{$\textit{R}$}}}A^{{\scalebox{0.64}{$\textit{R}$}}}_\mu+g_B^{{\scalebox{0.64}{$\textit{R}$}}}B^{{\scalebox{0.64}{$\textit{R}$}}}_\mu\right) \pi^{\mu \nu}\left(g_A^{{\scalebox{0.64}{$\textit{R}$}}}A^{{\scalebox{0.64}{$\textit{R}$}}}_\nu+g_B^{{\scalebox{0.64}{$\textit{R}$}}}B^{{\scalebox{0.64}{$\textit{R}$}}}_\nu\right)+{\gdualn{\mathfrak{f}}}^{{\scalebox{0.64}{$\textit{R}$}}}\upchi^{\mu \nu}{\mathfrak{f}}^{{\scalebox{0.64}{$\textit{R}$}}}\left(g_A^{{\scalebox{0.64}{$\textit{R}$}}}F^{{\scalebox{0.64}{$\textit{R}$}}}_{\mu \nu}+g_B^{{\scalebox{0.64}{$\textit{R}$}}}G^{{\scalebox{0.64}{$\textit{R}$}}}_{\mu \nu}\right)+ {\gdualn{\mathfrak{f}}}^{{\scalebox{0.64}{$\textit{R}$}}}\Sigma\ {\mathfrak{f}}^{{\scalebox{0.64}{$\textit{R}$}}}+\cdots,\nonumber \\
\eeq
being also implicit that there is an  infinite number of 1PI contributions to the quantum Lagrangian. Each bosonic external line carries the combination $(g_A^{{\scalebox{0.64}{$\textit{R}$}}}A_\mu^{{\scalebox{0.64}{$\textit{R}$}}}+g_B^{{\scalebox{0.64}{$\textit{R}$}}}B_\mu^{{\scalebox{0.64}{$\textit{R}$}}})$. 
 Here, according to the longitudinally projected Ward--Takahashi-like identities, $Z_MZ_B=1$ and $Z_AZ_\xi=1$, with the renormalized parameters being associated with the on-shell renormalization scheme. The bosonic wave function counter terms are employed to generate $\pi_{{\scalebox{0.64}{$\textit{R}$}}}^{\mu \nu}(0)=0$ to have a massless pole with residue equaling the unit for the physical sector in such a way to keep the combination $(g_A^{{\scalebox{0.64}{$\textit{R}$}}}A_\mu^{{\scalebox{0.64}{$\textit{R}$}}}+g_B^{{\scalebox{0.64}{$\textit{R}$}}}B_\mu^{{\scalebox{0.64}{$\textit{R}$}}})$ for the external lines.  To be possible to recover the standard Podolsky structure  even in the interacting case, the renormalized self-energy must be the same for the massless and the massive fields. Regarding the renormalization of the fermion sector, it is applied to yield the standard on-shell scheme. \\
 \indent The massive field renormalization with the presence of the factor $\mathfrak{F}_{{\scalebox{0.64}{$\textit{R}$}}}\equiv \frac{g_B^{{\scalebox{0.64}{$\textit{R}$}}}(\mu)}{g_A^{{\scalebox{0.64}{$\textit{R}$}}}(\mu)}$ is necessary to have a coupling for the massive field, which may receive different radiative corrections than the one associated to the massless field. 
Then, the effective massive propagator presents a normalization such that all the bosonic internal lines of 1PI functions are associated with the coupling $g_A^{{\scalebox{0.64}{$\textit{R}$}}}$, the physical one. The coupling $g_B^{{\scalebox{0.64}{$\textit{R}$}}}$ appears just when attached to the external massive bosonic lines. The function $\pi_{\mu \nu}=\theta_{\mu \nu}p^2\pi(p^2)$, whose transverse nature is ensured by Ward identities, denotes the photon self-energy without the coupling constants associated with the external bosonic lines.
The $\upchi^{\mu \nu}$ denotes the radiative corrections to the vertex functions defined up to the coupling associated with the external bosonic line. Its general form is obtained considering the vertex transversal nature. This is guaranteed by the Ward identity and implies that the electric and magnetic fields enter directly into the interaction. The ellipsis at the end denotes  the infinite sum of all classes of 1PI functions with their correlated external fields attached to it.\\
 \indent It is worth mentioning that just the $1$-loop photon and fermion self-energies are divergent, with all the remaining contributions being finite. The divergent piece of these objects can be obtained by the lowest-order iteration of the Schwinger--Dyson equations. Considering the previous definitions, they are expressed as
 \beq \mathcal{\pi}_{\mu \nu}(p)&=&4p^2\theta_{\mu \nu}\frac{\mu^{-\epsilon}}{16\pi^2}\left(  -\frac{2}{\epsilon} \right)\\
 \Sigma_{{\scalebox{0.6}{$\textsc{}$}}}^{{\scalebox{0.6}{$\textsc{}$}}}(p)&=&-3M^2_{{\scalebox{0.64}{$\textit{R}$}}}\ {\big(g_A^{{\scalebox{0.64}{$\textit{R}$}}}\big)}^2\frac{\mu^{-\epsilon}}{16\pi^2}\left(-\frac{2}{\epsilon} \right),  \eeq
 with the dimensional regularization parameter being defined in the limit $\epsilon \to 0$. 
 
 The knowledge of the explicit form of the only divergent graphs allied to its superrenormalizable nature allows us to conclude that the model is indeed renormalizable in such a way that all the physical on-shell conditions are capable of entirely fixing the counter terms and also eliminate all the $1/\epsilon$ singular terms.\\
 \indent The finite renormalization of the vertex  is chosen to keep the mentioned external field combination.\\  The counter terms are employed to furnish
 \beq\lim\limits_{\substack{
    p\to0\\
     q\to 0}}\upchi_{{\scalebox{0.64}{$\textit{R}$}}}^{\mu \nu}(p,q)=i(g^A_{{\scalebox{0.64}{$\textit{R}$}}})^2\sigma^{\mu \nu}\eeq
  with $q $ denoting the incoming DM momentum in the vertex diagrams and $p$ being the bosonic momentum. Owing to \cite{Ro}, and the transverse constraint on the vertex, this condition can indeed be fulfilled.  \\
 \indent Considering the second equivalent form for the vertex Schwinger--Dyson equation, at its lowest order, yields the finite object\footnote{The power on the coupling is following our previous discussion about the definition of the $\upchi_{\mu \nu}$ object.} to be renormalized
 \bea \upchi_{\mu \nu}(p,q)= {\big(g^A_{{\scalebox{0.64}{$\textit{R}$}}}\big)}^2M^2_{{\scalebox{0.64}{$\textit{R}$}}}\int \frac{d^4k }{(2\pi)^4} \frac{i\sigma^{\mu \nu}  }{[k^2-m^2_{{\scalebox{0.64}{$\textit{R}$}}}][(k+p)^2-m^2_{{\scalebox{0.64}{$\textit{R}$}}}][(k-q)^2-M^2_{{\scalebox{0.64}{$\textit{R}$}}}]}. \eea
 
After the operator mixing due to renormalization, the variations with respect to the massive field can be expressed as
\bea \frac{\delta}{\delta B_\mu}=\frac1{Z_B\mathfrak{F}_{{\scalebox{0.64}{$\textit{R}$}}}}\frac{\delta}{\delta B^{{\scalebox{0.64}{$\textit{R}$}}}_\mu}+\frac1{\tilde{\delta} }\theta_{\mu \nu}\frac{\delta}{\delta A^{{\scalebox{0.64}{$\textit{R}$}}}_\nu}\eea
 It explains why the Ward--Takahashi-like  identities, being related to longitudinal sectors, can just furnish constraints for the diagonal constants.\\
\indent The usual Podolsky model with mass $M_{{\scalebox{0.64}{$\textit{R}$}}}$ is recovered by
\bea A^{{\scalebox{0.64}{$\textit{R}$}}}_\mu(x)\mapsto A^{{\scalebox{0.64}{$\textit{R}$}}}_\mu(x)-\mathfrak{F}_{{\scalebox{0.64}{$\textit{R}$}}}\theta_\mu^{\ \nu}B^{{\scalebox{0.64}{$\textit{R}$}}}_\nu(x),\eea
followed by completing the square in $B_\mu^{{\scalebox{0.64}{$\textit{R}$}}}(x)$ field and eliminating a non-dynamical term from the quantum action. The renormalized quantum action in the standard Podolskyan form is given by
 \beq
    {\cal{L}}_{{\scalebox{0.6}{$\textsc{pod}$}}}^{{\scalebox{0.6}{$\textsc{quant}$}}}&=&-\frac{1}{4}F_{\mu \nu}^{{\scalebox{0.64}{$\textit{R}$}}}F^{\mu \nu}_{ {{\scalebox{0.64}{$\textit{R}$}}}}+\frac1{2M_{{\scalebox{0.64}{$\textit{R}$}}}^2}{\partial_\mu F^{\mu \nu}_{{\scalebox{0.64}{$\textit{R}$}}}\partial^\rho F_{\rho \nu}^{{\scalebox{0.64}{$\textit{R}$}}}}+\partial_\mu {\gdualn{\mathfrak{f}}}^{{\scalebox{0.64}{$\textit{R}$}}} \partial^\mu {\mathfrak{f}}^{{\scalebox{0.64}{$\textit{R}$}}}-m^2_{{\scalebox{0.64}{$\textit{R}$}}}\gdualn{\mathfrak{f}}^{{\scalebox{0.64}{$\textit{R}$}}} {\mathfrak{f}}^{{\scalebox{0.64}{$\textit{R}$}}}+\frac{ig_A^{{\scalebox{0.64}{$\textit{R}$}}}}{2}{\gdualn{\mathfrak{f}}}^{{\scalebox{0.64}{$\textit{R}$}}} \sigma_{\mu \nu} {\mathfrak{f}}^{{\scalebox{0.64}{$\textit{R}$}}} F^{\mu \nu}_{{\scalebox{0.64}{$\textit{R}$}}} \nonumber \\&& \qquad+\frac{1}{2}A^{{\scalebox{0.64}{$\textit{R}$}}}_\mu \Pi^{\mu \nu  }_{{\scalebox{0.64}{$\textit{R}$}}} A^{{\scalebox{0.64}{$\textit{R}$}}}_\nu+{\gdualn{\mathfrak{f}}}^{{\scalebox{0.64}{$\textit{R}$}}}g^{{\scalebox{0.64}{$\textit{R}$}}}_A\upchi^{\mu \nu }_{{\scalebox{0.64}{$\textit{R}$}}}{\mathfrak{f}}^{{\scalebox{0.64}{$\textit{R}$}}}F^{{\scalebox{0.64}{$\textit{R}$}}}_{\mu \nu}+{\gdualn{\mathfrak{f}}}^{{\scalebox{0.64}{$\textit{R}$}}}\Sigma_{{\scalebox{0.64}{$\textit{R}$}}}\ {\mathfrak{f}}^{{\scalebox{0.64}{$\textit{R}$}}}+\frac{\lambda_{{\scalebox{0.64}{$\textit{R}$}}}}{2}\left(\partial^\mu A_\mu^{{\scalebox{0.64}{$\textit{R}$}}} \right)^2+\cdots
\eeq
with $\Pi^{{\scalebox{0.64}{$\textit{R}$}}}(x-y)_{\mu \nu}$ denoting the standard form of the renormalized photonic self-energy tensor.\\
\indent The previous results guarantee that there is no need of adding counterterms not originally present in the bare Lagrangian to obtain a well-defined renormalized system. Hence, the Schwinger--Dyson equation determining the form of the fermion propagator can be employed without the need for modifications. Therefore, considering the discussion of the previous section, the renormalization properties are such that the Schwinger--Dyson equations do not need to be modified in any order. Then, it furnishes a reliable background for our discussion about the darkness mechanism.\\

 \section{concluding remarks and outlook}
\label{conclu}
The concept of darkness, naturally arising from \textcolor{black}{mass dimension one spinor fields}, was investigated, with the implementation of a Pauli-like interaction in the context of generalized electrodynamics, with a massive photon and a 
Stueckelberg field. The associated  Hamiltonian formalism was introduced and analyzed, using the Dirac--Bergmann algorithm. We, considering the Merlin mode concept, argued that the fermion-photon model asymptotic Hamiltonian operator has a positive definite spectrum, even though the fermionic sector has several quantum field-theoretical non-trivial features.

On constructing the path integral from the phase space structure with the  first and second-order Dirac constraints setup, the Ward-Takahashi-like identities and the quantum gauge symmetry were scrutinized in the context of \textcolor{black}{fermionic mass dimension one fields}, in which we show that no current-like vertex is formed due to radiative effects, for arbitrary orders. Appropriate generating functionals were engendered, in the path integral formalism, to evaluate  the vacuum-to-vacuum DM-photon scattering  amplitude, also using the Faddeev--Senjanovic Hamiltonian path-integral procedure. The  Schwinger--Dyson equations and the complete quantum equations were investigated throughout the text. In this context, the Schwinger-Dyson equations for the fermionic propagator were derived. These  equations for the photon propagator have been also derived as well as the one  for the vertex part. The vertex function was shown to depend on the 4-point fermion function. It is worth emphasizing that the possibility of two equivalent expressions for the vertex function yielded a non-perturbative relation between two whole classes of graphs composed of an infinite number
of diagrams, being similar  to the QED${}_4$ case, with the only difference due to  the presence of double
internal bosonic lines, denoting the superposition of massive and massless propagators, inherent to the model here used. We also obtained a nontrivial result associated with the form of the vertex until the $2$-loop order.

The Ward--Takahashi-like identities and the quantum gauge symmetry were investigated. 
The bare vertex was shown to be identically transverse, and the associated complete self-energy is gauge-invariant, differing  from the well-known standard QED${}_4$ case, where the fermionic self-energy is gauge-dependent and just the structure with attached external fermionic lines is gauge-independent. Another  relevant result consists of  considering full radiative corrections, yielding the fermions to couple solely to the
physical fields, with no direct coupling to the gauge field. On scrutinizing the non-polarized Compton-like and pair-annihilation processes, we noted that strong darkness, associated with vanishing amplitude due to the massless photon dispersion relation, is valid up to order $g^6$. We considered the experimental bounds in the model's parameters and also the bounds on the nucleon recoil cross-section. The analysis indicates that $g$ must be much smaller than the known standard model couplings, which means that the $g^6$ order is good enough for phenomenological darkness.

\indent Additionally, there is the definition of darkness in its weak form, associated with the vanishing amplitude for Compton and pair-annihilation unpolarized processes at $1$-loop on low scattering angles, whose 
probability vanishes. The weak darkness property is valid at least at order $g^8$. When considering the next first-order corrections to the Schwinger--Dyson
equation regulating the scattering amplitude, new classes of contributions can be considered. A particular case was addressed, involving the 4-point correlation box diagram associated
with the quantum action variations concerning two fermions and two photons. The darkness property was shown to be violated if an arbitrary number of loop corrections is considered, although, from
the four external field box, this prominent  feature still holds for the remaining 1-loop corrections. We discussed and stressed the fact that at cosmological scales, where low-energy
background photons and low momentum transfer are regarded, and both the tree level and the 1-loop properties ensure phenomenological darkness.

 To summarize, the Ward--Takahashi-like identities were studied for the DM-photon model and the propagators were computed, presenting a Merlin mode for the massive bosonic field. 
Therefore, the precise concept of  darkness was addressed for scattering amplitudes  in Compton-like processes involving dark matter and considering radiative corrections. The renormalizability 
of the theory, describing DM-generalized photon interactions, was also discussed and illustrated with Feynman diagrams. The renormalization procedure was thoroughly discussed, also taking into account a renormalized quantum Lagrangian with an infinite number of 1PI contributions.  These considerations were necessary to furnish a solid foundation for the formulation of darkness, since it involves radiative corrections and correlated subjects as well. We provided the explicit calculation of the divergent part of the bosonic and fermionic self-energies, as well as the vertex function at one loop. Since the model is superrenomalizable, these are the only divergent radiative corrections. The theory is finite beyond the 1-loop order.  \\
\indent Accomplishing our investigation, we demonstrated that a fermion-photon interaction is well-defined and under DM phenomenology. Therefore, considering the recent experimental research on DM-nucleon recoil processes, we derived a theory that allows these DM interactions with laboratory devices, without violating its defining property.

\subsubsection*{Acknowledgements}
GBG thanks to The São Paulo Research Foundation -- FAPESP  Post Doctoral grant No. 2021/12126-5. AAN thanks UNIFAL for its hospitality in his temporary stay as visiting professor.
RdR~is grateful to FAPESP (Grants No. 2021/01089-1 and No. 2022/01734-7) and the National Council for Scientific and Technological Development -- CNPq (Grant No. 303390/2019-0), for partial financial support.

\bibliography{ELKO.bib}

\providecommand{\newblock}{}
\begin{thebibliography}{10}
\expandafter\ifx\csname url\endcsname\relax
  \def\url#1{{\tt #1}}\fi
\expandafter\ifx\csname urlprefix\endcsname\relax\def\urlprefix{URL }\fi
\providecommand{\eprint}[2][]{\url{#2}}

\bibitem{Ahluwalia:2022ttu}
Ahluwalia D~V, da~Silva J~M~H, Lee C~Y, Liu Y~X, Pereira S~H and Sorkhi M~M
  2022 {\em Phys. Rept.\/} {\bf 967} 1--43 (\textit{Preprint}
  \eprint{2205.04754})

\bibitem{Ahluwalia:2004ab}
Ahluwalia D~V and Grumiller D 2005 {\em JCAP\/} {\bf 07} 012 (\textit{Preprint}
  \eprint{hep-th/0412080})

\bibitem{Lee:2015sqj}
Lee C~Y and Dias M 2016 {\em Phys. Rev. D\/} {\bf 94} 065020 (\textit{Preprint}
  \eprint{1511.01160})

\bibitem{Alves:2014kta}
Alves A, de~Campos F, Dias M and Hoff~da Silva J~M 2015 {\em Int. J. Mod. Phys.
  A\/} {\bf 30} 1550006 (\textit{Preprint} \eprint{1401.1127})

\bibitem{Dias:2010aa}
Dias M, de~Campos F and Hoff~da Silva J~M 2012 {\em Phys. Lett. B\/} {\bf 706}
  352--359 (\textit{Preprint} \eprint{1012.4642})

\bibitem{Agarwal:2014oaa}
Agarwal B, Jain P, Mitra S, Nayak A~C and Verma R~K 2015 {\em Phys. Rev. D\/}
  {\bf 92} 075027 (\textit{Preprint} \eprint{1407.0797})

\bibitem{Duarte:2017svd}
Duarte L~C, Lima R~d~C, Rogerio R~J~B and Villalobos C~H~C 2019 {\em Adv. Appl.
  Clifford Algebras\/} {\bf 29} 66 (\textit{Preprint} \eprint{1705.10302})

\bibitem{Alves:2017joy}
Alves A, Dias M, de~Campos F, Duarte L and Hoff~da Silva J~M 2018 {\em EPL\/}
  {\bf 121} 31001 (\textit{Preprint} \eprint{1712.05180})

\bibitem{Alves:2014qua}
Alves A, Dias M and de~Campos F 2014 {\em Int. J. Mod. Phys. D\/} {\bf 23}
  1444005 (\textit{Preprint} \eprint{1410.3766})

\bibitem{XENON:2019ykp}
Aprile E {\em et~al.\/} (XENON) 2019 {\em Phys. Rev. D\/} {\bf 100} 052014
  (\textit{Preprint} \eprint{1906.04717})

\bibitem{Aprile:2022vux}
Aprile E {\em et~al.\/} 2022  (\textit{Preprint} \eprint{2207.11330})

\bibitem{Sassi:2022njl}
Sassi S, Heikinheimo M, Tuominen K, Kuronen A, Byggm\"astar J, Nordlund K and
  Mirabolfathi N 2022 {\em Phys. Rev. D\/} {\bf 106} 063012 (\textit{Preprint}
  \eprint{2206.06772})

\bibitem{daRocha:2011yr}
da~Rocha R, Bernardini A~E and Hoff~da Silva J~M 2011 {\em JHEP\/} {\bf 04} 110
  (\textit{Preprint} \eprint{1103.4759})

\bibitem{daRocha:2009gb}
da~Rocha R and Hoff~da Silva J~M 2009 {\em Int. J. Geom. Meth. Mod. Phys.\/}
  {\bf 6} 461--477 (\textit{Preprint} \eprint{0901.0883})

\bibitem{deBrito:2019hih}
de~Brito G~P, Hoff Da~Silva J~M and Nikoofard V 2020 {\em Eur. Phys. J. ST\/}
  {\bf 229} 2023--2034 (\textit{Preprint} \eprint{1912.02912})

\bibitem{Bonora:2014dfa}
Bonora L, de~Brito K~P~S and da~Rocha R 2015 {\em JHEP\/} {\bf 02} 069
  (\textit{Preprint} \eprint{1411.1590})

\bibitem{Pereira:2017efk}
Pereira S~H and Guimar\~aes T~M 2017 {\em JCAP\/} {\bf 09} 038
  (\textit{Preprint} \eprint{1702.07385})

\bibitem{Rodrigues:2005yz}
Rodrigues Jr W~A, da~Rocha R and Vaz Jr J 2005 {\em Int. J. Geom. Meth. Mod.
  Phys.\/} {\bf 2} 305 (\textit{Preprint} \eprint{math-ph/0501064})

\bibitem{Ahluwalia:2004sz}
Ahluwalia D~V and Grumiller D 2005 {\em Phys. Rev. D\/} {\bf 72} 067701
  (\textit{Preprint} \eprint{hep-th/0410192})

\bibitem{Pereira:2021dkn}
Pereira S~H 2022 {\em Int. J. Mod. Phys. D\/} {\bf 31} 2250056
  (\textit{Preprint} \eprint{2110.12890})

\bibitem{XENON:2020rca}
Aprile E {\em et~al.\/} (XENON) 2020 {\em Phys. Rev. D\/} {\bf 102} 072004
  (\textit{Preprint} \eprint{2006.09721})

\bibitem{Ahluwalia:2022yvk}
Ahluwalia D~V, da~Silva J~M~H and Lee C~Y 2023 {\em Nucl. Phys. B\/} {\bf 987}
  116092 (\textit{Preprint} \eprint{2212.13114})

\bibitem{deGracia:2022enm}
de~Gracia G~B and da~Rocha R 2022  (\textit{Preprint} \eprint{2206.11989})

\bibitem{Podolsky:1942zz}
Podolsky B 1942 {\em Phys. Rev.\/} {\bf 62} 68--71

\bibitem{Podolsky:1944zz}
Podolsky B and Kikuchi C 1944 {\em Phys. Rev.\/} {\bf 65} 228--235

\bibitem{Bertin:2009gs}
Bertin M~C, Pimentel B~M and Zambrano G~E~R 2011 {\em J. Math. Phys.\/} {\bf
  52} 102902 (\textit{Preprint} \eprint{0907.1078})

\bibitem{Galvao:1986yq}
Galvao C~A~P and Pimentel~Escobar B~M 1988 {\em Can. J. Phys.\/} {\bf 66}
  460--466

\bibitem{Cuzinatto:2005zr}
Cuzinatto R~R, de~Melo C~A~M and Pompeia P~J 2007 {\em Annals Phys.\/} {\bf
  322} 1211--1232 (\textit{Preprint} \eprint{hep-th/0502052})

\bibitem{Bufalo:2010sb}
Bufalo R, Pimentel B~M and Zambrano G~E~R 2011 {\em Phys. Rev. D\/} {\bf 83}
  045007 (\textit{Preprint} \eprint{1008.3181})

\bibitem{El-Bennich:2020aiq}
El-Bennich B, Ramos-Zambrano G and Rojas E 2021 {\em Phys. Rev. D\/} {\bf 103}
  076008 (\textit{Preprint} \eprint{2010.15993})

\bibitem{Donoghue:2021eto}
Donoghue J~F and Menezes G 2021 {\em Phys. Rev. D\/} {\bf 104} 045010
  (\textit{Preprint} \eprint{2105.00898})

\bibitem{Faddeev:1969su}
Faddeev L~D 1969 {\em Theor. Math. Phys.\/} {\bf 1} 1--13

\bibitem{Senjanovic:1976br}
Senjanovic P 1976 {\em Annals Phys.\/} {\bf 100} 227--261 [Erratum: Annals
  Phys. 209, 248 (1991)]

\bibitem{Dirac:1958jc}
Dirac P~A~M 1959 {\em Phys. Rev.\/} {\bf 114} 924--930

\bibitem{Arina:2020mxo}
Arina C, Cheek A, Mimasu K and Pagani L 2021 {\em Eur. Phys. J. C\/} {\bf 81}
  223 (\textit{Preprint} \eprint{2005.12789})

\bibitem{Ahluwalia:2019etz}
Ahluwalia D 2019 {\em {Mass Dimension One Fermions}\/} vol 229 (Cambridge
  University Press) (\textit{Preprint} \eprint{2007.15098})

\bibitem{Speranca:2013hqa}
Speran\c{c}a L~D 2014 {\em Int. J. Mod. Phys. D\/} {\bf 23} 1444003
  (\textit{Preprint} \eprint{1304.4794})

\bibitem{daRocha:2008we}
da~Rocha R and Hoff~da Silva J~M 2010 {\em Adv. Appl. Clifford Algebras\/} {\bf
  20} 847--870 (\textit{Preprint} \eprint{0811.2717})

\bibitem{Ablamowicz:2014rpa}
Ab\l{}amowicz R, Gon\c{c}alves I and da~Rocha R 2014 {\em J. Math. Phys.\/}
  {\bf 55} 103501 (\textit{Preprint} \eprint{1409.4550})

\bibitem{daRocha:2007sd}
da~Rocha R and Pereira J~G 2007 {\em Int. J. Mod. Phys. D\/} {\bf 16}
  1653--1667 (\textit{Preprint} \eprint{gr-qc/0703076})

\bibitem{Fabbri:2016msm}
Fabbri L 2016 {\em Int. J. Geom. Meth. Mod. Phys.\/} {\bf 13} 1650078
  (\textit{Preprint} \eprint{1603.02554})

\bibitem{Ahluwalia:2008xi}
Ahluwalia D~V, Lee C~Y and Schritt D 2010 {\em Phys. Lett. B\/} {\bf 687}
  248--252 (\textit{Preprint} \eprint{0804.1854})

\bibitem{Wigner:1939cj}
Wigner E~P 1939 {\em Annals Math.\/} {\bf 40} 149--204

\bibitem{HoffdaSilva:2019eao}
Hoff~da Silva J~M and Bueno~Rogerio R~J 2019 {\em EPL\/} {\bf 128} 11002
  (\textit{Preprint} \eprint{1908.00458})

\bibitem{Podolsky:1945chv}
Podolsky B and Kikuchi C 1945 {\em Phys. Rev.\/} {\bf 67} 184

\bibitem{Bufalo:2014jra}
Bufalo R, Pimentel B~M and Soto D~E 2014 {\em Phys. Rev. D\/} {\bf 90} 085012
  (\textit{Preprint} \eprint{1407.1476})

\bibitem{Ruegg:2003ps}
Ruegg H and Ruiz-Altaba M 2004 {\em Int. J. Mod. Phys. A\/} {\bf 19} 3265--3348
  (\textit{Preprint} \eprint{hep-th/0304245})

\bibitem{m2y}
Bonin C~A, de~Gracia G~B, Nogueira A~A and Pimentel B~M 2022 {\em Braz. J.
  Phys.\/} {\bf 52} 127 (\textit{Preprint} \eprint{1911.01127})

\bibitem{Ro}
K\i{}z\i{}lers\"u A, Sizer T, Pennington M~R, Williams A~G and Williams R 2015
  {\em Phys. Rev. D\/} {\bf 91} 065015 (\textit{Preprint} \eprint{1409.5979})

\bibitem{Robert1}
Roberts C~D 2008 {\em Prog. Part. Nucl. Phys.\/} {\bf 61} 50--65

\bibitem{Robert2}
Bashir A, Chang L, Cloet I~C, El-Bennich B, Liu Y~X, Roberts C~D and Tandy P~C
  2012 {\em Commun. Theor. Phys.\/} {\bf 58} 79--134 (\textit{Preprint}
  \eprint{1201.3366})

\bibitem{Albino}
L~Albino A~Bashir A~M and Raya A 2022  (\textit{Preprint}
  \eprint{hep-ph/2210.01280})

\bibitem{Nash}
Nash C 2011 {\em {Relativistic Quantum Fields}\/}

\bibitem{Landau}
Landau and Lifshitz 1982 {\em {Quantum electrodynamics}\/}

\bibitem{Peskin:1995ev}
Peskin M~E and Schroeder D~V 1995 {\em {An Introduction to quantum field
  theory}\/} (Reading: Addison-Wesley)

\bibitem{Duarte:2020svn}
Duarte L, Dias M and de~Campos F 2020 {\em Eur. Phys. J. ST\/} {\bf 229}
  2133--2146

\bibitem{LUX:2016ggv}
Akerib D~S {\em et~al.\/} (LUX) 2017 {\em Phys. Rev. Lett.\/} {\bf 118} 021303
  (\textit{Preprint} \eprint{1608.07648})

\bibitem{XENON:2017vdw}
Aprile E {\em et~al.\/} (XENON) 2017 {\em Phys. Rev. Lett.\/} {\bf 119} 181301
  (\textit{Preprint} \eprint{1705.06655})

\bibitem{LZ:2022ufs}
Aalbers J {\em et~al.\/} (LZ) 2022  (\textit{Preprint} \eprint{2207.03764})

\bibitem{Aalbers:2022dzr}
Aalbers J {\em et~al.\/} 2022  (\textit{Preprint} \eprint{2203.02309})

\bibitem{David}
d’Enterria D and da~Silveira G~G 2013 {\em Phys. Rev. Lett.\/} {\bf 111}
  080405

\bibitem{Martellini:1997mu}
Martellini M and Zeni M 1997 {\em Phys. Lett. B\/} {\bf 401} 62--68
  (\textit{Preprint} \eprint{hep-th/9702035})

\end{thebibliography}

\end{document}